\documentclass[%
 reprint,
 amsmath,amssymb,
 aps,
pra,
]{revtex4-1}

\usepackage{graphicx}
\usepackage{dcolumn}
\usepackage{bm}

\usepackage{amsmath}
\usepackage{amssymb}
\usepackage{graphicx,xcolor}
\usepackage{subfigure}
\usepackage{adjustbox}
\usepackage{multirow}

\usepackage{epstopdf}

\newcommand{\be}[1]{\begin{equation}\label{#1}}
\newcommand{\ee}{\end{equation}}
\newcommand{\ba}[1]{\begin{eqnarray}\label{#1}}
\newcommand{\ea}{\end{eqnarray}}
\newcommand{\rf}[1]{(\ref{#1})}
\newcommand{\nn}{\nonumber}

\begin{document}


\title{Detecting singular weak-dissipation limit for flutter onset in reversible systems}

\author{Davide Bigoni}
\affiliation{University of Trento, via Mesiano 77, 38123 Trento, Italy}
\author{Oleg N. Kirillov}%
\email{Oleg.Kirillov@northumbria.ac.uk}
\affiliation{Northumbria University, Newcastle upon Tyne, NE1 8ST, UK}
\author{Diego Misseroni}
\affiliation{University of Trento, via Mesiano 77, 38123 Trento, Italy}
\author{Giovanni Noselli}
\affiliation{SISSA-International School for Advanced Studies, via Bonomea 265, 34136 Trieste, Italy}
\author{Mirko Tommasini}
\affiliation{University of Trento, via Mesiano 77, 38123 Trento, Italy}


\date{\today}

\begin{abstract}

{A `flutter machine' is introduced for the} investigation of a singular interface between the classical and reversible Hopf bifurcations that is theoretically predicted to be generic in nonconservative reversible systems with vanishing dissipation. In particular, such a singular interface exists for the Pfl\"uger viscoelastic column moving in a resistive medium, which is proven by means of the perturbation theory of multiple eigenvalues with the Jordan block. The laboratory setup, consisting of a cantilevered viscoelastic rod loaded by a positional force with non-zero curl produced by dry friction, demonstrates high sensitivity of the classical Hopf bifurcation onset {to the ratio between} the weak air drag and Kelvin-Voigt damping in the Pfl\"uger column. Thus,
the Whitney umbrella singularity is experimentally confirmed, responsible for discontinuities accompanying dissipation-induced instabilities in a broad range of physical contexts.
\end{abstract}

\maketitle


\section{\label{sec:level1}Introduction}

In a dissipative system oscillatory flutter instability, an example of a classical Hopf bifurcation, shifts a complex-conjugate pair of eigenvalues to the right in the complex plane. This instability mechanism is modified for a non-dissipative system possessing a \textit{reversible} symmetry,
defined with reference to the differential equation
$$\frac{d {\bf x}}{d t}={\bf g}({\bf x}),\quad {\bf x}\in \mathbb{R}^n$$
{which is said to be} $\bf R$-reversible (${\bf R}^{-1}={\bf R}$) if it is invariant with respect to the transformation $({\bf x}, t) \mapsto ({\bf R}{\bf x}, -t)$, implying that the right hand side {must} satisfy ${\bf R}{\bf g}({\bf x})=- {\bf g} ({\bf R}{\bf x})$.

If ${\bf x}={\bf x}_0$ is a reversible equilibrium such that ${\bf R}{\bf x}_0={\bf x}_0$, and ${\bf A}=\nabla {\bf g}$ is the
{linearization matrix about ${\bf x}_0$,} then ${\bf A}=-{\bf R}{\bf A}{\bf R}$, and the characteristic polynomial $$\det({\bf A}-\lambda {\bf I})=\det(-{\bf R}{\bf A}{\bf R}-{\bf R}\lambda {\bf R})=(-1)^n\det({\bf A}+\lambda {\bf I}),$$ {implies} that $\pm\lambda, \pm\overline{\lambda}$ are eigenvalues of ${\bf A}$ \cite{OR1996,W2003,Clerc1999,Clerc2001}.
Due to the {spectrum's symmetry} with respect to both the real and imaginary axes of the complex plane, the reversible-Hopf bifurcation requires {the generation} of a non-semi-simple double pair of imaginary eigenvalues and its subsequent separation into a complex quadruplet \cite{OR1996,W2003,Clerc1999,Clerc2001}.

{All equations} of second order
$$\frac{d^2 {\bf x}}{d t^2}={\bf f}({\bf x}),$$
{are reversible \cite{OR1996,W2003}, including the case} when
the positional force ${\bf f}({\bf x})$ has a non-trivial curl, $\nabla \times {\bf f}({\bf x}) \ne 0$, which makes the reversible system \textit{nonconservative}.

Such nonconservative \textit{curl forces} \cite{BS2012} appear in modern opto-mechanical applications, including optical tweezers  \cite{G2003,Wu2009,SH2010}. In mechanics, they are known as \textit{circulatory forces} for producing non-zero work along a closed circuit. Circulatory forces are common in the models of friction-induced vibrations \cite{HG2003}, rotordynamics \cite{Clerc2001},  biomechanics \cite{BD2016} and fluid-structure interactions \cite{MM2010,G2017}, to name a few. A circulatory force acting on an elastic structure and remaining directed along the tangent line to the structure at the point of its application during deformation is known as \textit{follower} \cite{Z1952,B1963,KS1998}.

Since the dynamics of an elastic structure under a follower load is described by reversible equations \cite{OR1996}, flutter instability may occur via the reversible-Hopf bifurcation mechanism \cite{OR1996,Clerc2001}. In these conditions, Ziegler \cite{Z1952} discovered that, when viscosity is present, the location of the curve for the onset of the classical Hopf bifurcation is displaced by an order-one distance in the parameter space, with respect to the curve for the onset of the reversible-Hopf bifurcation in the elastic structure. This occurs  even if the viscous damping in the structure is infinitesimally small \cite{Z1952}. Other velocity-dependent forces, such as air drag (or even gyroscopic forces), can also destabilize an elastic structure under a follower load \cite{OR1996,SW1975,T2016,K2013dg,KV2010}. However, acting together, the velocity-dependent forces, e.g.,
the air drag and the material (Kelvin-Voigt) viscous damping, can \textit{inhibit} the destabilizing effect of each other at a particular \textit{ratio} of their magnitudes due to the singular interface between the classical Hopf and reversible-Hopf bifurcations \cite{B1956,Z1952,KV2010,T2016}.


For instance, the system
\be{example}
\ddot {\bf x}(t)+(\delta {\bf D} + \Omega {\bf G})\dot {\bf x}(t) + ({\bf K} +\nu {\bf N}){\bf x}(t)=0, \quad {\bf x} \in \mathbb{R}^2
\ee
{where $\delta$, $\Omega$, $\nu$ are scalar coefficients and matrices ${\bf D}>0$, ${\bf K}>0$ are real and symmetric, while matrices {\bf G} and {\bf N} are
skew-symmetric as follows}
$${\bf G}={\bf N}=\left(
\begin{array}{rr}
0 & -1 \\
1 & 0 \\
\end{array}
\right),
$$
is nonconservative and reversible for $\delta=\Omega=0$.

The reversible-Hopf bifurcation in the system \rf{example} with $\delta=\Omega=0$ occurs at
$$\nu_f=\sqrt{\omega_f^4-\det {\bf K}},\quad \omega_f^2=\frac{{\rm tr} {\bf K}}{2},$$ where ``$\rm tr$'' denotes the trace {operator, which yields flutter
instability when} $\nu>\nu_f$. However, {when} $\delta>0$, $\Omega>0$ the classical-Hopf bifurcation occurs at a different value of $\nu$ \cite{K2013dg}
$$
\nu_H(\Omega,\delta)\approx\nu_f-\frac{2\nu_f}{({\rm tr} {\bf D})^2}\left(\frac{\Omega}{\delta}-\frac{{\rm tr}({\bf KD}-\omega_f^2{\bf D})}{2\nu_f}\right)^2.
$$

The expression for $\nu_H(\Omega,\delta)$ defines a surface in the $(\delta,\Omega,\nu)$-space that has a Whitney's umbrella singular point at $(0,0,\nu_f)$ \cite{L2003,footnote}. Near that singular point, the neutral stability surface is a ruled surface, with a self-intersection degenerating at the singularity, so that {a unique} value of the ratio $\Omega/\delta$
is produced, for which the onsets of the classical and reversible Hopf bifurcations tend to coincide \cite{B1956,KV2010,footnote}.

For a dissipative
{nearly-reversible} system, the singular dependence of the classical Hopf bifurcation onset on {the} parameters of velocity-dependent forces has a general character \cite{B1956}, which
follows from the codimension 3 (for dissipative systems) and 1 (for reversible vector fields) of non-semi-simple double imaginary eigenvalues \cite{B1956, A1972, KV2010,KH2010, K2013dg}.

Since the singularity is related to a  double imaginary eigenvalue arising from a Jordan block \cite{A1972},
it can be found in other dissipative systems that are close to undamped systems with the `reversible' symmetry of spectrum  \cite{KV2010}.

Indeed, the system \rf{example} with $\delta=0$, $\Omega=0$, and $\nu=0$ is a conservative Hamiltonian system, which is statically unstable for ${\bf K}<0$.
Adding gyroscopic forces with $\Omega>0$, keeps this system Hamiltonian and yields its stabilization if $\Omega> \Omega_f=\sqrt{-\kappa_1}+\sqrt{-\kappa_2}$, where $\kappa_{1,2}<0$ are eigenvalues of $\bf K$. Owing to the `reversible' symmetry of its spectrum \cite{OR1996,MK1991,BKMR94}, the Hamiltonian system displays flutter instability via the collision of imaginary eigenvalues at $\Omega=\Omega_f$ and their subsequent splitting into a complex quadruplet as soon as $\Omega$ decreases below $\Omega_f$.
This is the {so-called} linear Hamilton-Hopf bifurcation \cite{L2003,K2013dg,K2007}.

If $\delta>0$, $\nu>0$ the gyroscopic stability is destroyed at the threshold of the classical-Hopf bifurcation \cite{K2013dg,K2007}
$$
\Omega_H\approx\Omega_f+\frac{2 \Omega_f}{(\omega_f {\rm tr} {\bf D})^2}\left(\frac{\nu}{\delta} - \frac{{\rm tr}({\bf KD}+(\Omega_f^2-\omega_f^2){\bf D})}{2\Omega_f}\right)^2,
$$
where $\omega_f^2=\sqrt{\kappa_1\kappa_2}$ and ${\bf D}>0$. The dependency of the {new gyroscopic stabilization threshold} just on the ratio $\nu/\delta$ implies that the limit of $\Omega_H$ as
{both $\nu$ and $\delta \rightarrow 0$} is higher than $\Omega_f$  for all ratios except a unique one. Similarly to the case of nonconservative reversible systems, this happens because the classical Hopf and the Hamilton-Hopf bifurcations meet in the Whitney umbrella singularity that exists on the stability boundary of a
{nearly-Hamiltonian} dissipative system and corresponds to the onset of the Hamilton-Hopf bifurcation \cite{L2003,KH2010,K2007,KV2010,K2013dg,BKMR94,KM2007}.

The singular weak-dissipation limit for the flutter onset in {nearly-Hamiltonian} systems in the presence of two different damping mechanisms
has been discovered first in the problem of secular instability of equilibria of rotating and self-gravitating masses of fluid, when dissipation due to both fluid viscosity \cite{TT1879,RS1963,BO2014} and emission of gravitational waves \cite{C1970,C1984} is taken into account \cite{LD1977,A2003}. Later on this phenomenon manifested itself as the \lq Holop\"ainen instability mechanism' for a baroclinic flow \cite{H1961,WE2012} and as an enhancement of modulation instability with dissipation \cite{BD2007}. Analysis of this effect based on the method of normal forms and perturbation of multiple eigenvalues has been
developed, {among others by} \cite{B1956,OR1996,AY1974,A1972,KH2010,KI2005,K2013dg,K2007,BKMR94,Clerc2001,KI2017,LF2016,KS2005,MK1991,KV2010,FS1978,S1980}.

                  \begin{figure}
    \begin{center}
    \includegraphics[angle=0, width=\columnwidth]{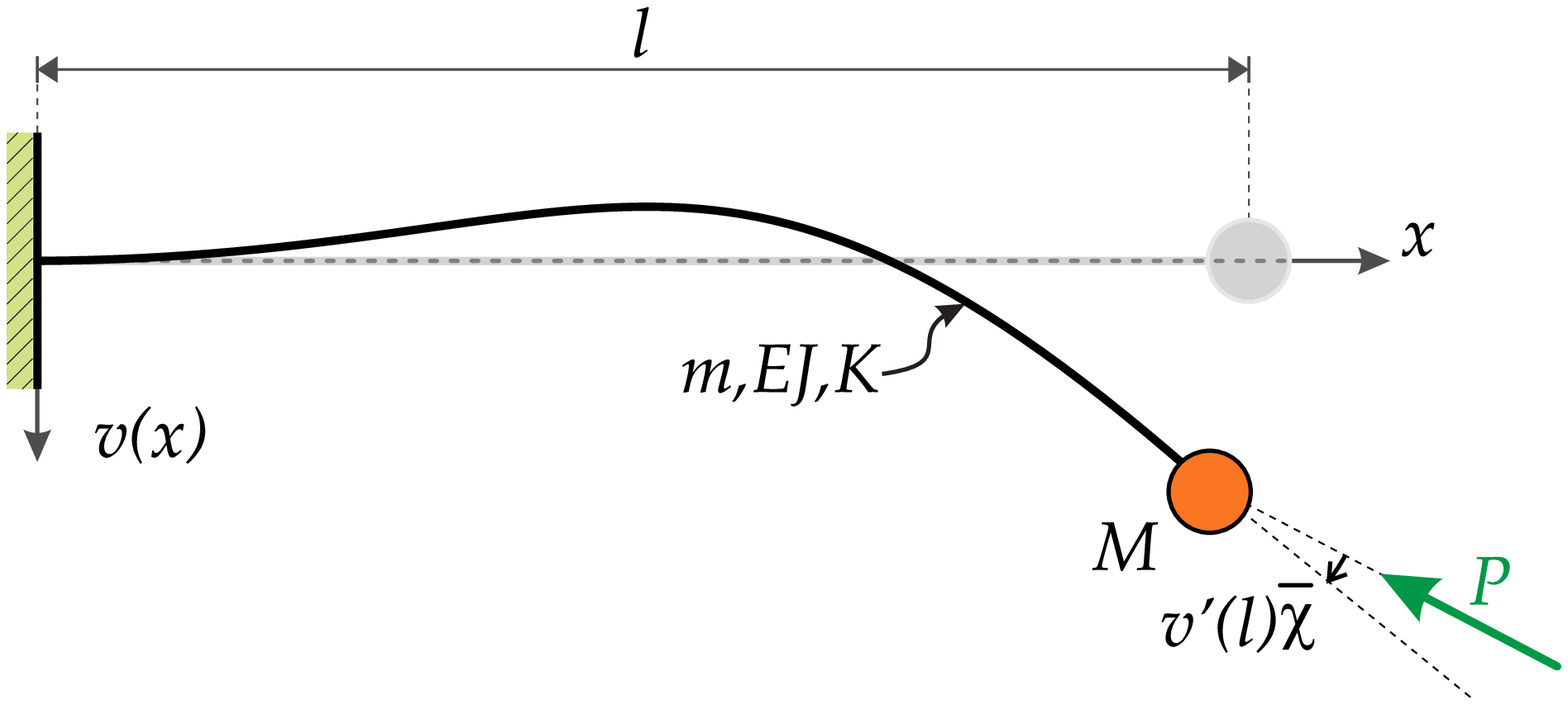}
    \end{center}
    \caption{The  Pfl\"uger column \cite{P1955} clamped at $x=0$ with a point mass $M$ at $x=l$. The column is loaded at $x=l$
		{with} a constant compressing circulatory force $P$ inclined to the tangent to the elastic line of the column, so that $v'(l)\bar \chi = const.$ (equal to 0.092 in all the experiments).  }
    \label{fig1}
    \end{figure}

Although the destabilizing effect of damping for equilibria of Hamiltonian and reversible systems has been discussed for decades, no experimental evidence is known
for the singular limit of the classical Hopf bifurcation in a {nearly-Hamiltonian}, or a {nearly-reversible} system, when
{the} dissipation tends to zero. The main difficulty for such experiments is the accurate identification and control of at least two different damping mechanisms. For reversible elastic structures an additional challenge lies in the realization of circulatory follower loads, acting for a sufficiently long time. Previous attempts are reported to create a follower load through the thrust produced either by water flowing through a nozzle \cite{WSS1969}, or by a solid rocket motor mounted at the end of an elastic rod in a cantilever configuration \cite{SKK1995,SMR1995,SKK2000,LS2000}. In the former  realization hydrodynamical effects  enter into play and in the latter the duration of the experiments is limited to a few seconds. In contrast, the frictional follower force acting on a wheel mounted at the free end of the double-link Ziegler pendulum allowed Bigoni and Noselli to significantly relax the limitation on time \cite{BN2011}.

In the present article, an experimental realization is reported {for} the  Pfl\"uger column \cite{B1952,P1955,
SKK1976,RS2003},  a viscoelastic cantilevered rod carrying a point mass at the free end and loaded
{with} a follower force (Fig.~\ref{fig1}) obtained via friction, similarly to \cite{BN2011}. Two {dissipation} mechanisms---the air resistance and the internal Kelvin-Voigt damping---are identified and controlled by changing the geometrical characteristics of the sample rods. The measured critical flutter loads demonstrate a high sensitivity to the ratio between the two damping coefficients, being almost insensitive to each of the damping coefficients that both are very close to zero, in agreement with both numerical modeling of \cite{T2016} and perturbation theory developed for the Pfl\"uger column in the present work.


\section{\label{sec:level1}Pfl\"uger's column and its Galerkin discretization}

Consider a rod of length $l$, mass density per unit length $m$ and end mass $M${, its deflection $v$, function of the $x$ coordinate, obeys the Bernoulli law that the rotation of the {cross-}section $\phi$ is given by}
$\phi(x) = -v'(x)$, where a prime denotes derivative with respect to $x$. A {moment}-curvature viscoelastic constitutive relation of the Kelvin-Voigt type is assumed in the form
$$
\mathcal{M}(x,t) = - EJ v''(x,t) - E^*J \dot{v}''(x,t),
$$
where a superimposed dot denotes the time derivative, $E$ and $E^*$ are respectively the elastic and the viscous moduli of the rod, which has a cross section with moment of inertia $J$. The rod is clamped at one end and is loaded {through} the force $P$ that is inclined with respect to the tangent to the rod at its free end such that $v'(l)\bar \chi=const.$, Fig.~\ref{fig1}.

Assuming that a distributed external damping $K$ caused by the air drag is acting on the rod, and introducing the dimensionless quantities
\ba{dimensionless}
&\xi =\frac{x}{l}, ~ \tau = \frac{t}{l^2}\sqrt{\frac{E J}{m}}, ~ p =\frac{P l^2}{E J}, ~\alpha =\tan^{-1}\left(\frac{M}{m l}\right),& \nn\\
&\eta = \frac{E^*l^2}{\sqrt{m E J}}{\frac{ {J}}{l^4}}, ~\gamma = \frac{K l^2}{\sqrt{m EJ}}, ~\beta=\frac{\gamma}{\eta},~\chi=1-\bar{\chi},&
\ea
the linearized partial differential equation of motion governing the dynamics of the rod can be written as
\ba{}
\label{checazzo}
&v''''(\xi, \tau) + \eta \dot{v}''''(\xi,\tau) + p v''(\xi,\tau) + \gamma \dot{v}(\xi,\tau) + \ddot{v}(\xi,\tau) = 0,&\nn\\
&~&
\ea
where now a prime and a dot denote partial differentiation with respect to $\xi$ and $\tau$, respectively.
Separating time in (\ref{checazzo}) with
$
v(\xi,\tau)=\tilde{v}(\xi) \exp(\omega \tau)
$
yields a non-self-adjoint boundary eigenvalue problem \cite{T2016}
\ba{beckeq}
&(1+ \eta\omega) \tilde{v}'''' + p \tilde{v}'' +  (\gamma \omega  + \omega^2)\tilde{v} = 0,&\nn\\
&(1+\eta \omega)\tilde{v}'''(1) - (\chi-1)\tilde{v}'(1)p-\omega^2 \tan(\alpha) \tilde{v}(1) =0,&\nn\\
&\tilde{v}(0)=\tilde{v}'(0) = 0,\quad \tilde{v}''(1) = 0.&
\ea
Assuming that $\tilde{v}(\xi)$ has the form
\ba{hyperbolic_f}
\tilde{v}(\xi)&=A_1 \sinh(\lambda_1 \xi) + A_2 \cosh(\lambda_1 \xi)&\nn\\
 &+ A_3 \sin (\lambda_2 \xi) + A_4 \cos(\lambda_2 \xi),&
\ea
with $A_i$ ($i=1,..,4$) arbitrary constants and
\be{char_eq}
\lambda^2_{1,2}=\frac{\sqrt{p^2-4(1+\eta \omega)(\gamma \omega+\omega^2)} \mp p}{2(1+\eta \omega)}
\ee
and substituting (\ref{hyperbolic_f}) into (\ref{beckeq}) yields an algebraic system of equations which admits non-trivial solutions if \cite{T2016}
\ba{trans_eq}
&0=\lambda_1\lambda_2(1+\eta \omega)(\lambda_1^4+\lambda_2^4)+\lambda_1\lambda_2 p(\chi-1)(\lambda_2^2-\lambda_1^2)& \nn\\
&+\lambda_1\lambda_2[2(1{+}\eta \omega)\lambda_1^2\lambda_2^2{-}p(\chi{-}1)(\lambda_2^2{-}\lambda_1^2)]\cosh \lambda_1 \cos \lambda_2&\nn \\
&-\omega^2 \tan \alpha (\lambda_1^2{+}\lambda_2^2)[\lambda_2\sinh\lambda_1\cos\lambda_2{-}\lambda_1\cosh\lambda_1\sin\lambda_2]&\nn\\
&+\lambda_1^2 \lambda_2^2[2p(\chi-1){+}(1+\eta \omega)(\lambda_2^2{-}\lambda_1^2)]\sinh \lambda_1\sin \lambda_2.&
\ea

{
Results from experiments are compared} with the eigenvalues, eigenfunctions and critical parameters of the boundary eigenvalue problem \rf{beckeq} which
{are directly found} by numerical solution of the transcendental characteristic equation \rf{trans_eq}.

For theoretical purposes{, the $N$-dimensional Galerkin discretization of the continuous problem \rf{beckeq} is also considered}:
\ba{eip}
(\omega^2[{\bf I}{+}4{\bf M}_1\tan \alpha]{+}\omega[\gamma{\bf I}{+}\eta{\bf D}_i]
{+}\left[{\bf K}_1{-}p{\bf K}_2{+}\chi p{\bf N}\right]){\bf a}{=}0,\nn\\
\ea
where $\bf a$ is an $N$-vector and $\bf I$ is the $N\times N$ identity matrix. The entries of the $N\times N$ mass matrix ${\bf M}_1$ are $M_{1,ij}=(-1)^{i+j}$, the matrix of internal damping ${\bf D}_i$ is ${\bf D}_i={\rm diag}(\omega_1^2,\omega_2^2,\ldots,\omega_N^2)$, and the stiffness matrix ${\bf K}_1$ is ${\bf K}_1={\rm diag}(\omega_1^2,\omega_2^2,\ldots,\omega_N^2)$. The values of the frequencies $\omega_1,\ldots,\omega_N$ as well as the entries of the symmetric stiffness matrix ${\bf K}_2$ and the non-symmetric matrix of circulatory forces $\bf N$ are given in the Appendix \ref{App A}.

\section{Theory of dissipation-induced flutter instability}\label{pt}

For a Galerkin-discretized model of the Pfl\"uger column \rf{eip} {a perturbation theory is developed} of the singular weak-dissipation limit for the onset of flutter.

\subsection{The $N=2$ modes approximation and its stability analysis}\label{SectionPT4}

          \begin{figure*}
    \begin{center}
    \includegraphics[angle=0, width=\textwidth]{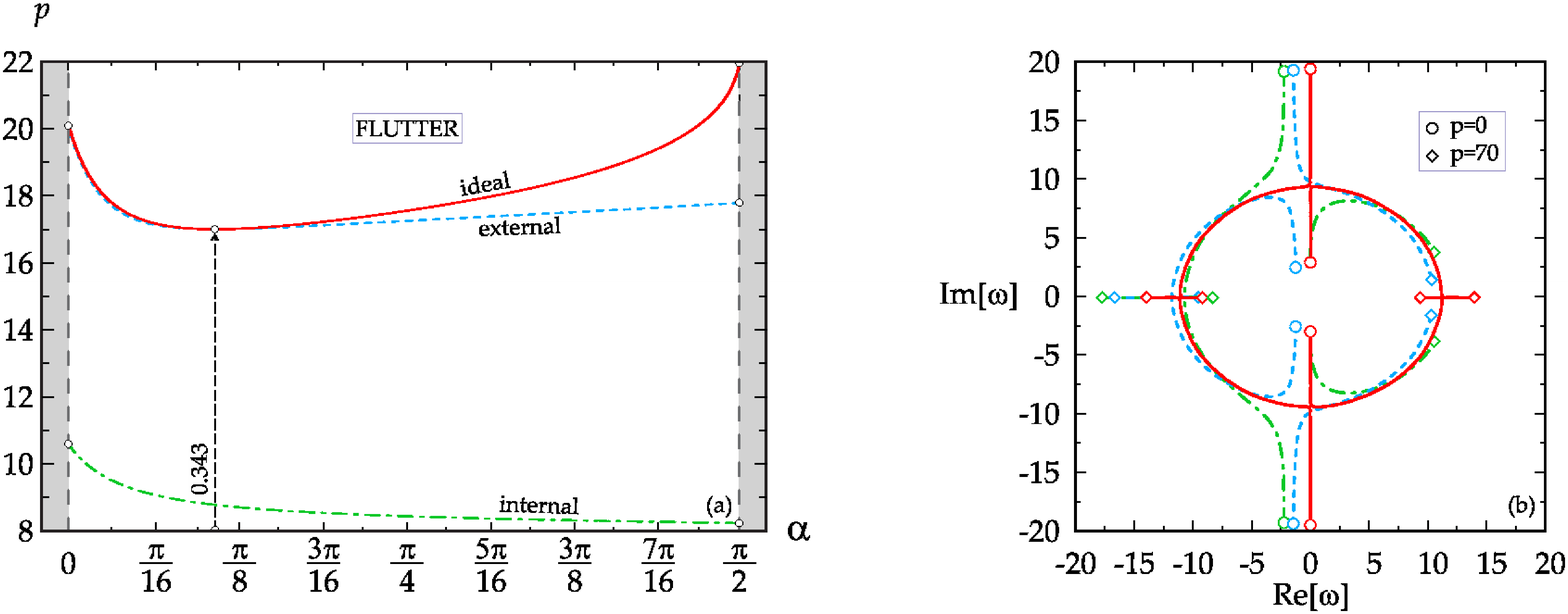}
    \end{center}
    \caption{(a) Stability boundary for (green dash-dot curve) internally and (blue dashed curve) externally damped discretized model of the Pfl\"uger column with $N=2$ modes and $\chi=1$, when one of the damping coefficients is zero and another one tends to zero. The red solid curve shows {the} stability boundary of the non-damped discretized model of the Pfl\"uger column according to Eq.~\rf{flubou}. (b) The eigenvalue movement when $p$ {increases} from $0$ (circle) to $70$ (diamond)
	for $N=2$, $\chi=1$, $\alpha=0.1$, and 	(red solid curves) $\gamma=0$, $\eta=0$, (blue dashed curves) $\gamma=4.5$, $\eta=0$, and (green dash-dotted curves) $\gamma=0$, $\eta=0.015$.}
    \label{figp1}
    \end{figure*}

    The eigenvalue problem \rf{eip} has the form
\be{eip2}
({\bf M}(\alpha)\omega^2+{\bf D}(\gamma,\eta)\omega+{\bf A}(p,\chi)){\bf a}=0,
\ee
where ${\bf M}={\bf M}^T$, ${\bf D}={\bf D}^T$, ${\bf D}(0,0)=0$, and ${\bf A}\ne {\bf A}^T$, with the superscript $T$ denoting transposition.
\\\\
Recall that the adjugate ${\bf X}^{*}$ of a $N\times N$ matrix $\bf X$ is defined as ${\bf X}^{*}={\bf X}^{-1}\det{\bf X}$ and, in particular,
\be{adjm}
 \frac{\partial \det {\bf X}}{\partial p}={\rm tr}\left({\bf X}^* \frac{\partial \bf X}{\partial p}\right).
\ee
{Since ${\rm tr}({\bf X}^*{\bf Y})={\rm tr}({\bf Y}^*{\bf X})$ for $N=2$, the} characteristic polynomial of \rf{eip2} in the case of $N=2$  can be written by means of the Leverrier algorithm in a compact
form:
\ba{charp2}
&q(\omega,\alpha,\chi,p,\gamma,\eta)=\det{\bf M}\omega^4+{\rm tr}({\bf D}^{*}{\bf M})\omega^3&\nn\\
&+({\rm tr}({\bf A}^{*}{\bf M})+\det{\bf D})\omega^2+{\rm tr}({\bf A}^*{\bf D})\omega+\det{\bf A}.&
\ea
{Assuming} that for $\eta=0$, $\gamma=0$, $\alpha=\alpha_0$, $\chi=\chi_0$, and $p=p_0$ the undamped system with $N=2$ degrees of freedom be on the flutter boundary, {on this boundary} its eigenvalues are imaginary and form a double complex-conjugate pair $\omega=\pm i \sigma_0$ of a Jordan block.
In these conditions, the real critical frequency $\sigma_0$ at the onset of flutter follows from the characteristic polynomial in the closed form
\ba{freq2}
&\sigma_0^2=\frac{{\rm tr}({\bf A}_0^{*}{\bf M}_0)}{2\det{\bf M}_0}=\sqrt{\frac{\det {\bf A}_0}{\det {\bf M}_0}},&\nn\\
&{\bf M}_0={\bf M}(\alpha_0),\quad {\bf A}_0={\bf A}(p_0,\chi_0)&
\ea
and the flutter boundary is {described} by the equation
\be{flubou}
({\rm tr}({\bf A}_0^{*}{\bf M}_0))^2=4\det{{\bf A}_0}\det{{\bf M}_0}.
\ee
Since ${\bf M}_0={\bf I}+4{\bf M}_1\tan \alpha_0$ and ${\bf A}_0={\bf K}_1-p_0{\bf K}_2+\chi_0 p_0{\bf N}$ is a linear function of $p_0$, {equation \rf{flubou} is quadratic} with respect to $p_0$,
which can thus be easily solved. The red solid curve in Fig.~\ref{figp1}(a) shows the flutter boundary \rf{flubou} of the undamped discretized model \rf{eip} of the Pfl\"uger column with $N=2$ modes for $\chi_0=1$ in the $(\alpha_0,p_0)$-plane. The red solid curves in Fig.~\ref{figp1}(b) demonstrate {the movement of the}  eigenvalues of the undamped system at given $\chi=\chi_0=1$ and $\alpha=\alpha_0=0.1$ when the load parameter $0\le p\le 70$. The equilibrium is stable for $0\le p < p_0$ where the critical flutter load is $p_0\approx17.83368$, corresponding to a double pair of imaginary eigenvalues with the imaginary part $\sigma_0\approx9.366049$ (see Eq.~\rf{freq2}).  The value $p=p_0$ corresponds to the linear reversible-Hopf bifurcation, yielding {the} splitting of the double eigenvalues into a complex quadruplet causing flutter instability.
\subsection{Reversible-Hopf bifurcation in the undamped model}\label{SectionPT5}

{A perturbation formula is now derived} for the splitting of a double eigenvalue $\omega=i\sigma_0$, when $\gamma=\gamma_0$ and $\alpha=\alpha_0$ are fixed and $p$ {is left to} vary. Introducing a small parameter $0\le\varepsilon \ll 1$ and assuming in the polynomial $q_0(\omega,p)=q(\omega,\alpha_0,\chi_0,p,\gamma=0,\eta=0)$ that $p(\varepsilon)=p_0+\varepsilon \frac{d p}{d \varepsilon}+\ldots$ {(where the derivative is taken at $\varepsilon=0$) yields}
\ba{expol}
& q_0(\omega,p(\varepsilon))=\sum_{r=0}^{2N}\frac{(\omega(\varepsilon)-i\sigma_0)^r}{r!}\left(\frac{\partial^r q_0}{\partial \omega^r} +\varepsilon \frac{\partial^r q_1}{\partial \omega^r}+o(\varepsilon)\right),&\nn\\
& \frac{\partial^r q_1}{\partial \omega^r}=\frac{\partial^{r+1}q_0}{\partial \omega^r\partial p}\frac{d p}{d \varepsilon},&
\ea
where the partial derivatives are evaluated at $p=p_0$ and $\omega=i\sigma_0$.

Assuming for the perturbed double non-semisimple eigenvalue the Newton-Puiseux series
\be{npe}
\omega(\varepsilon)=i\sigma_0+\varepsilon^{1/2}\sigma_1+\varepsilon\sigma_2+\ldots,
\ee
substituting  equations \rf{expol} and \rf{npe} into the equation $q_0(\omega,p)$ and collecting the terms of the same powers of $\varepsilon$,
{leads to}
\be{doube}
q_0(i\sigma_0,p_0)=0,\quad
\left.\sigma_1\frac{\partial q_0}{\partial \omega}\right|_{\omega=i\sigma_0,p=p_0}=0,
\ee
and
\be{doub2}
\left.\left(q_1+\frac{1}{2}\sigma_1^2\frac{\partial^2 q_0}{\partial \omega^2}+\sigma_2\frac{\partial q_0}{\partial \omega}\right)\right|_{\omega=i\sigma_0,p=p_0}=0.
\ee
Conditions \rf{doube} are satisfied for the double eigenvalue $\omega=i\sigma_0${, so that an account of this into \rf{doub2} yields}
$$
\sigma_1^2=-q_1\left(\frac{1}{2}\frac{\partial^2 q_0}{\partial \omega^2} \right)^{-1}=-\left(\frac{1}{2}\frac{\partial^2 q_0}{\partial \omega^2} \right)^{-1} \frac{\partial^{}q_0}{\partial p}\frac{d p}{d \varepsilon}.
$$
Hence, the splitting of the double non-semisimple eigenvalue due to {the} variation of $p$ is governed by the formula
$$
\omega(p)=i\sigma_0 \pm i\sqrt{\left(\frac{1}{2}\frac{\partial^2 q_0}{\partial \omega^2} \right)^{-1} \frac{\partial q_0}{\partial p}(p-p_0)} + o(|p-p_0|^{1/2}).
$$
With the help of Eq.~\rf{adjm}, Eq.~\rf{freq2}, and the relations
\ba{relat}
&q_0(\omega,p)=\omega^4\det{\bf M}+\omega^2{\rm tr}({\bf M}^{*}{\bf A})+\det{\bf A},&\nn\\
&\left.\frac{\partial q_0}{\partial p}\right|_{\omega=i\sigma_0,p=p_0}=-{\rm tr}\left(({\bf A}_0^*-\sigma_0^2{\bf M}_0^{*})({\bf K}_2-\chi_0 {\bf N})\right),&\nn\\
&\left.\frac{1}{2}\frac{\partial^2 q_0}{\partial \omega^2}\right|_{\omega=i\sigma_0,p=p_0}=-2{\rm tr}({\bf A}_0^{*}{\bf M}_0),&
\ea
{the following result is finally obtained}
\ba{traf}
&\omega(p)=i\sigma_0&\\
& \pm i\sqrt{\frac{{\rm tr}\left[({\bf A}_0^*-\sigma_0^2{\bf M}_0^{*})({\bf K}_2-\chi_0 {\bf N})\right]}{2{\rm tr}({\bf A}_0^{*}{\bf M}_0)}(p-p_0)} + o(|p-p_0|^{1/2}).&\nn
\ea
For instance, for $\alpha_0=0.1$, $\chi_0=1$, $p_0\approx17.83368$, and  $\sigma_0\approx9.366049$, the expression \rf{traf} {becomes}
\be{traf1}
\omega(p)\approx i\sigma_0 \pm i\sqrt{-3.962532(p-p_0)}
\ee
confirming the splitting of the double $i\sigma_0$ into two complex eigenvalues with opposite real parts (flutter) at $p>p_0$.

\subsection{Dissipative perturbation of simple imaginary eigenvalues}\label{SectionPT6}

At $p<p_0$ the eigenvalues of the undamped system $\omega=\omega(p)$ remain simple and imaginary. To investigate how they are affected by dissipation,
{it is assumed that}  $\eta(\varepsilon)=\frac{d \eta}{d\varepsilon}\varepsilon+o(\varepsilon)$, and $\gamma(\varepsilon)=\frac{d \gamma}{d\varepsilon}\varepsilon +o(\varepsilon)$ in the polynomial \rf{charp2}, where {$\alpha=\alpha_0$, $\gamma=\gamma_0$ and $0\le p<p_0$ are also fixed}.
Then, $\omega = \omega(p)+\frac{d \omega}{d \varepsilon} \varepsilon+o(\varepsilon)$, with
$$
\frac{d \omega}{d \varepsilon}=-\left(\frac{\partial q}{\partial \omega} \right)^{-1}\left(
\frac{\partial q}{\partial \eta}\frac{d \eta}{d \varepsilon}+\frac{\partial q}{\partial \gamma}\frac{d \gamma}{d \varepsilon}\right).
$$
{The following approximation is therefore obtained}
$$
\omega=\omega(p)-\left(\frac{\partial q}{\partial \omega} \right)^{-1}\left(
\frac{\partial q}{\partial \eta}\eta+\frac{\partial q}{\partial \gamma}\gamma\right)+o(\gamma,\eta) ,
$$
where the partial derivatives are evaluated  at $p<p_0$ and $\omega=\omega(p)$. {An account of the following derivatives}
\ba{deri}
&\frac{\partial q}{\partial \omega}=2\sigma_0^{-2}\omega{\rm tr}({\bf M}_0^{*}(\omega^2{\bf A}_0+\sigma_0^2{\bf A})),&\nn\\
&\frac{\partial q}{\partial \eta}=\omega{\rm tr}\left({\bf D}_i^*({\bf A}+\omega^2{\bf M}_0)\right),&\nn\\
&\frac{\partial q}{\partial \gamma}=\omega{\rm tr}\left({\bf A}+\omega^2{\bf M}_0\right),&
\ea
{leads to}
\ba{omat}
\omega&=&\omega(p)\nn\\
&-&\frac{
\eta{\rm tr}\left({\bf D}_i^*({\bf A}+\omega^2{\bf M}_0)\right)+\gamma{\rm tr}\left({\bf A}+\omega^2{\bf M}_0\right)}{2{\rm tr}({\bf M}_0^{*}(\omega^2{\bf A}_0+\sigma_0^2{\bf A}))}\sigma_0^{2}\nn\\
&+&o(\gamma,\eta).
\ea

\subsection{Linear approximation to the stability boundary and the exact zero-dissipation limit of the critical flutter load}\label{SectionPT7}

\begin{figure*}
    \begin{center}
    \includegraphics[angle=0, width=\textwidth]{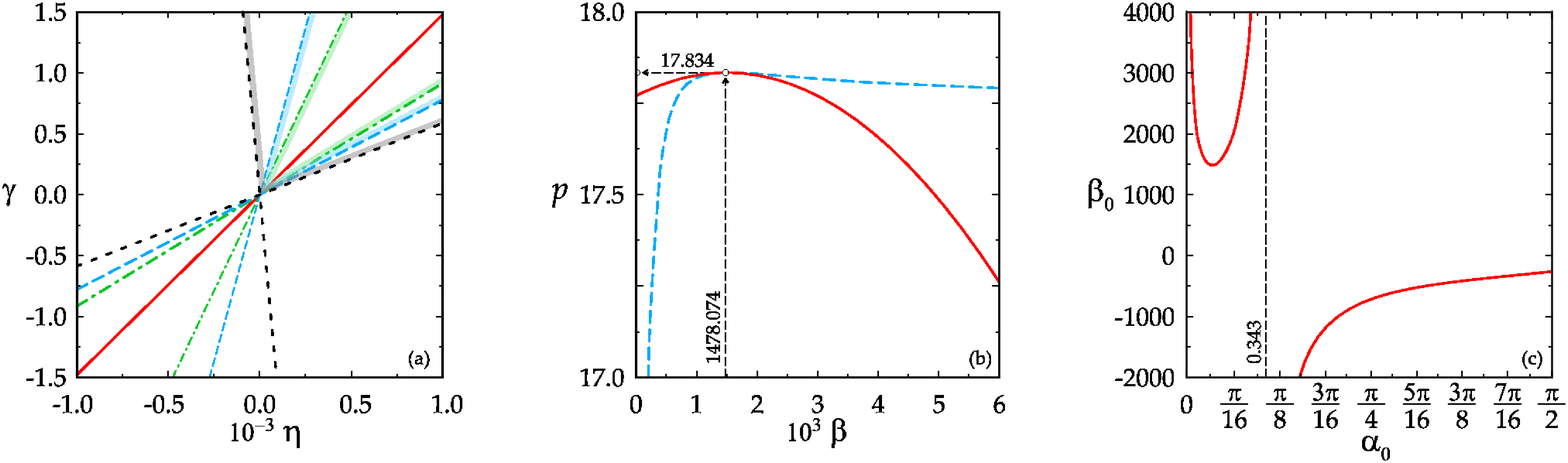}
      \end{center}
    \caption{(a) For $N=2$, $\chi_0=1$, and $\alpha_0=0.1$ the linear approximation \rf{wu1} to the classical-Hopf bifurcation onset in the $(\eta,\gamma)$-plane for (black dotted line) $p=p_0-0.1$, (blue dashed line) $p=p_0-0.04$, (green dot-dashed line) $p=p_0-0.02$, and (red solid line) $p=p_0$. {The stability region for} every $p$ is inside the narrow angle-shaped regions in the first quadrant; flutter instability in the complement. (b) The critical flutter load in the limit of vanishing dissipation as a function of the damping ratio $\beta=\gamma/\eta$ according to the (blue dashed curve) exact expression \rf{wu1} and (red solid curve) its quadratic approximation \rf{pbeta}. The maximum of the limit coincides with the critical flutter load $p_0\approx17.83368$ of the undamped system at $\beta=\beta_0\approx1478.074$ that is determined from Eq.~\rf{beta0}. (c) The stabilizing ratio $\beta_0$ as a function of $\alpha_0$ according to equation \rf{beta0} with vertical asymptotes at $\alpha_0=0$ (Beck's column) and $\alpha_0\approx 0.342716$.}
    \label{figp3}
    \end{figure*}

The {correction, linear in $\eta$ and $\gamma$,} to the simple imaginary eigenvalue in \rf{omat} due to damping is real and therefore it determines whether the dissipative perturbation is stabilizing or destabilizing.
Equating this linear term to zero
and taking into account that ${\bf A}={\bf K}_1-p({\bf K}_2-\chi_0 {\bf N})$ and ${\bf D}_i={\bf K}_1={\rm diag}(\omega_1^2,\omega_2^2)$
yields the following approximation to the flutter boundary, which represents the onset of the classical Hopf bifurcation
\ba{wu1}
&\eta\left[2\omega_1^2\omega_2^2+{\rm tr}(
{\bf D}_i^*({\bf M}_0\omega^2(p)-p({\bf K}_2-\chi_0 {\bf N})))\right]=&\nn\\
&-\gamma\left[\omega_1^2+\omega_2^2+{\rm tr}({\bf M}_0\omega^2(p)-p({\bf K}_2-\chi_0 {\bf N})\right],&
\ea
where ${\bf M}_0={\bf I}+4{\bf M}_1\tan \alpha_0$ and $\omega(p)$ is a root of the polynomial $q_0(\omega,p)$ in equation \rf{relat}$_1$ at $p<p_0$. In the $(\eta,\gamma)$-plane the equation \rf{wu1} defines a straight line, Fig.~\ref{figp3}(a). In fact, at every $p<p_0$ there exist two lines \rf{wu1}
corresponding to two different eigenvalues $\omega(p)$ that participate in the reversible-Hopf bifurcation at $p=p_0$. However, as $p$ tends to $p_0$, the angle between the two lines
decreases
and completely vanishes in the limit $p\rightarrow p_0$, Fig.~\ref{figp3}(a). This suggests that the approximation \rf{wu1} defines a ruled surface in the $(\eta,\gamma,p)$-space. As a consequence, every fixed damping ratio $\beta=\gamma/\eta$ corresponds to a ruler at some $p<p_0$. Therefore,
the condition for which the damping tends to zero at fixed damping ratio
will occur along this ruler for the corresponding constant value of $p<p_0$ and will result in the limiting value of the critical flutter load that is lower than the critical load at the onset of the reversible-Hopf bifurcation, $p_0$, see Fig.~\ref{figp3}(b). Note that equation \rf{wu1} gives the exact dependency of the limit of the critical flutter load at vanishing dissipation as a function of the damping ratio, $\beta$, {if the exact solution $\omega(p)$ of the polynomial $q_0(\omega,p)$ is used}, see \cite{K2013dg,KV2010,AY1974,K2007,LF2016}.
              \begin{figure*}
    \begin{center}
    \includegraphics[angle=0, width=\textwidth]{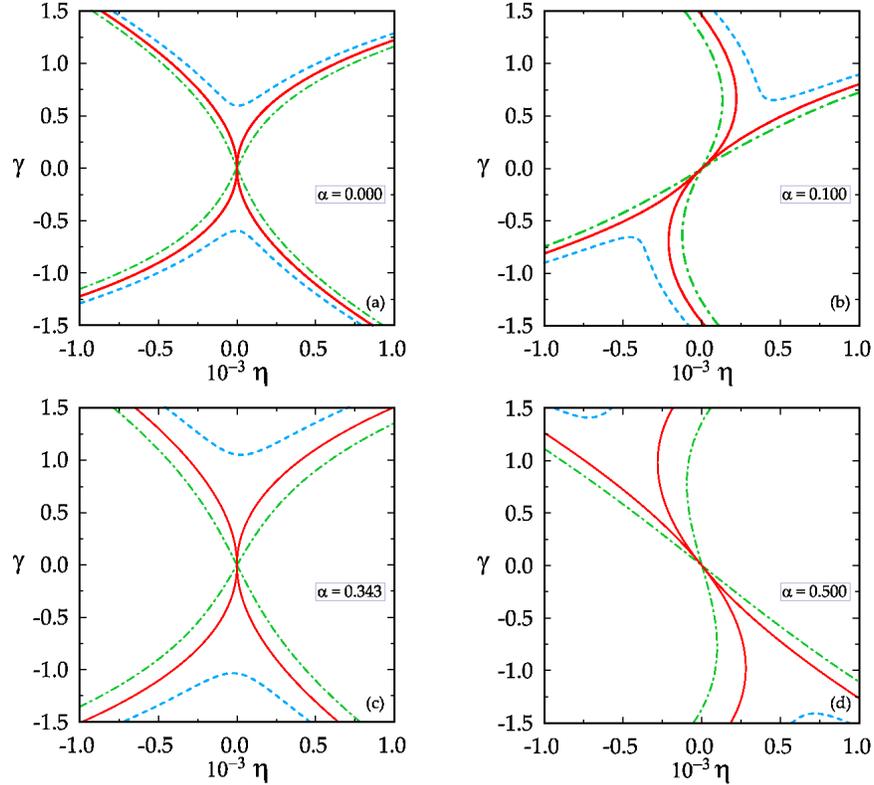}
    \end{center}
    \caption{For $N=2$, $\chi_0=1$, stability boundary of the discretized model for the Pfl\"uger column in the plane of internal, $\eta$, and external, $\gamma$, damping for (a) $\alpha_0=0$ with $\beta_0\rightarrow +\infty$, (b) $\alpha_0=0.1$ with $\beta_0\approx1478.074$, (c) $\alpha_0 \approx 0.3427$ with $\beta_0\rightarrow +\infty$, (d) $\alpha_0=0.5$ with $\beta_0\approx-1856.099$. The red solid lines correspond to the undamped critical load $p=p_0(\alpha_0)$, which depends on $\alpha_0$, the blue dashed lines to $p=p_0(\alpha_0)+0.02$, and the green dash-dotted lines to $p=p_0(\alpha_0)-0.02$.}
    \label{figp2}
    \end{figure*}

\subsection{Quadratic approximation in $\beta$ to the exact zero-dissipation limit of the critical flutter load}\label{SectionPT8}

In the vicinity of $p=p_0$ the two roots participating in the reversible-Hopf bifurcation are approximated by the expression \rf{traf}.
Using {this expression in equation} \rf{wu1}, {the limit of zero dissipation can be found} for the critical flutter load as a function of the damping ratio, $p(\beta)$, in the form of a series
\begin{widetext}
\be{pbeta}
p(\beta)=p_0-\frac{2{\rm tr}({\bf A}_0^{*}{\bf M}_0)}{{\rm tr}\left[({\bf A}_0^*-\sigma_0^2{\bf M}_0^{*})({\bf K}_2-\chi_0 {\bf N})\right]}\left[\frac{{\rm tr}\left({\bf A}_0-\sigma_0^2{\bf M}_0\right)}{2\sigma_0{\rm tr}({\bf M}_0^*(\beta_0{\bf I}+{\bf D}_i))}\right]^2(\beta-\beta_0)^2+o((\beta-\beta_0)^2),
\ee
\end{widetext}
where
\be{beta0}
\beta_0 = -\frac{{\rm tr}\left({\bf D}_i^*({\bf A}_0-\sigma_0^2{\bf M}_0)\right)}{{\rm tr}\left({\bf A}_0-\sigma_0^2{\bf M}_0\right)}.
\ee
From the quadratic approximation \rf{pbeta} it is evident that $p(\beta)\le p_0$ for all $\beta$ except for the specific case of $\beta=\beta_0$, at which it  exactly coincides with the critical flutter load of the undamped system: $p(\beta_0)=p_0$. For instance, for $\alpha_0=0.1$ and $\chi_0=1$, the approximation \rf{pbeta} is
\be{apwu}
p(\beta)\approx 17.83368-2.807584\cdot 10^{-8}(\beta-1478.074)^2,
\ee
as shown in Fig.~\ref{figp3}(b) with a red solid curve.

\subsection{The Whitney umbrella singularity}\label{SectionPT9}

Truncating the series \rf{pbeta} and substituting $\beta=\gamma/\eta$ into the result, {yields} an expression for the ruled surface in the $(\eta,\gamma,p)$-space
\begin{widetext}
\be{wu2}
p(\gamma,\eta)=p_0-\frac{2{\rm tr}({\bf A}_0^{*}{\bf M}_0)}{{\rm tr}\left[({\bf A}_0^*-\sigma_0^2{\bf M}_0^{*})({\bf K}_2-\chi_0 {\bf N})\right]}\left[\frac{{\rm tr}\left({\bf A}_0-\sigma_0^2{\bf M}_0\right)}{2\sigma_0{\rm tr}({\bf M}_0^*(\beta_0{\bf I}+{\bf D}_i))}\right]^2\frac{(\gamma-\beta_0\eta)^2}{\eta^2}.
\ee
\end{widetext}
This expression is in the form $Z=X^2/Y^2$, which is the well-known normal form for the Whitney umbrella surface [12-15]. The surface \rf{wu2} has a singular point at $p=p_0$, corresponding to the onset of the reversible-Hopf bifurcation, and a self-intersection at $p<p_0$.

In Fig.~\ref{figp2}
the cross-sections are plotted in the $(\eta,\gamma)$-plane for different values of $p$ of the exact stability boundary  calculated with the use of the Routh-Hurwitz criterion applied directly to the polynomial \rf{charp2}. Physically relevant is the first quadrant of the $(\eta,\gamma)$-plane.

For every $\alpha_0 \in [0,\pi/2]$ the cross-sections look qualitatively similar. For $p>p_0$ the stability domain is bounded by a smooth curve departing from the origin, Fig.~\ref{figp2}. For $p=p_0(\alpha_0)$ the stability boundary has a cuspidal point at the origin with the single tangent line to the boundary specified by the ratio $\beta_0$ given by the equation \rf{beta0}; {the stability region} is inside the cusp. For $p<p_0(\alpha_0)$ the stability boundary has a point of intersection at the origin in the $(\eta,\gamma)$-plane; {the stability region} is inside the narrow angle-shaped domain, which becomes wider as $p$ decreases and for $p=0$ spreads over the first quadrant of the plane for every possible mass distribution.

{A comparison between Fig.~\ref{figp3}(a) and Fig.~\ref{figp2}(b) shows} that equation \rf{wu1} gives a correct linear approximation to the stability domain provided by the Routh-Hurwith criterion in the $(\eta,\gamma)$-plane and, therefore, to the singular interface between the classical-Hopf and reversible-Hopf bifurcations in the $(\eta,\gamma,p)$-space.

\subsection{Stabilizing damping ratio $\beta_0$ for different mass distributions $\alpha_0$}\label{SectionPT10}

Fig.~\ref{figp2} demonstrates that the contour plot patterns of the stability boundary in the $(\eta,\gamma)$-plane {remain} qualitatively the same for different values of $\alpha_0${, but} differ in the orientation of the cusp, which is determined by the stabilizing damping ratio $\beta_0$. Evaluating \rf{beta0} at the points of the stability boundary of the undamped system,
{provides the plot of the function $\beta_0(\alpha_0)$ reported} in Fig.~\ref{figp3}(c). One can see that two intervals of $\alpha_0$ exist with opposite signs of $\beta_0$.  The intervals are bounded by the {values $\alpha_0=0$ and $\alpha_0\approx0.342716$,} at which the graph $\beta_0(\alpha_0)$ displays a vertical asymptote, Fig.~\ref{figp3}(c). Positive values of $\beta_0$ correspond to sufficiently small $\alpha_0\le0.342716$, cf. Fig.~\ref{figp2}(b); negative values of $\beta_0$ are characteristic for $0.342716\le\alpha_0\le \pi/2$.

{The above} critical values of $\alpha_0$ are determined by the zeros of the denominator of equation \rf{beta0}. Indeed, taking into account that
\ba{}
&{\rm tr}{\bf M}_0=2+8\tan \alpha_0,\quad {\rm det}{\bf M}_0=1+8\tan \alpha_0,&\nn\\
& {\rm tr}({\bf A}_0^{*}{\bf M}_0)=
{\rm tr}{\bf A}_0+4{\rm tr}({\bf M}_1^{*}{\bf A}_0)\tan \alpha_0,&
\ea
{the denominator can be obtained in the form}
\ba{}
&{\rm tr}({\bf A}_0-\sigma_0^2 {\bf M}_0)={\rm tr}{\bf A}_0-\frac{{\rm tr}({\bf A}_0^{*}{\bf M}_0)}{2\det{\bf M}_0}{\rm tr}{\bf M}_0&\nn\\
&=
\frac{4\tan \alpha_0}{1+8\tan \alpha_0}{\rm tr}\left(({\bf I}-(1+4\tan \alpha_0){\bf M}_1^*){\bf A}_0\right).&
\ea
Evidently, one of the roots {is $\alpha_0=0$}, corresponding to the case of the Beck column {(which is the Pfl\"uger column without the end mass)}. In this case, the cusp in the $(\eta,\gamma)$-plane is oriented vertically, see Fig.~\ref{figp2}(a). This confirms the well-known fact that for the Beck column the internal Kelvin-Voigt damping $(\eta)$ is destabilizing, and the external air drag damping $(\gamma)$ is stabilizing \cite{KS2005,B1963,T2016}. As soon as $\alpha_0$ departs from zero, the external damping becomes a destabilizing factor due to the change in the orientation of the cusp in Fig.~\ref{figp2}. Nevertheless, at a specific mass distribution $\alpha_0\approx0.342716$, which is given by the root of the equation
$$
{\rm tr}\left[({\bf I}-(1+4\tan \alpha_0){\bf M}_1^*){\bf A}_0\right]=0,
$$
the cusp restores its vertical orientation, as is visible in Fig.~\ref{figp2}(c). For this specific mass ratio the external damping is stabilizing again.

The revealed behaviour of the stabilizing damping ratio as a function of the mass distribution is reflected in Fig.~\ref{figp1}(a) that shows the red solid curve of the onset of the reversible-Hopf bifurcation in the undamped system together with the onset of the classical-Hopf bifurcation in the limit of vanishing (the green dash-dotted curve) internal damping and (the blue dashed curve) external damping. The latter curve has two common points with the stability boundary of the undamped system exactly at $\alpha_0=0$ and $\alpha_0\approx0.342716$.

Remarkably, $\beta_0$ and its sign determine which mode will be destabilized by either of the two damping mechanisms or by their combination. For instance, in the case of $\beta_0>0$ the cusp of the stability boundary in the $(\eta,\gamma)$-plane is directed to the first quadrant, Fig.~\ref{figp2}(b). Therefore, a dominating external damping will destabilize the mode with the higher frequency, whereas
a dominating internal damping will destabilize the mode with the lower frequency, see Fig.~\ref{figp1}(b). In the case of $\beta_0<0$ the cusp is oriented towards the second quadrant, Fig.~\ref{figp2}(d),
so that for every choice of internal and external damping with $\eta>0$ and $\gamma>0$, the mode with the lower frequency will be the destabilizing one.

\begin{figure}[bt!]
  \begin{center}
     \includegraphics[width=\columnwidth]{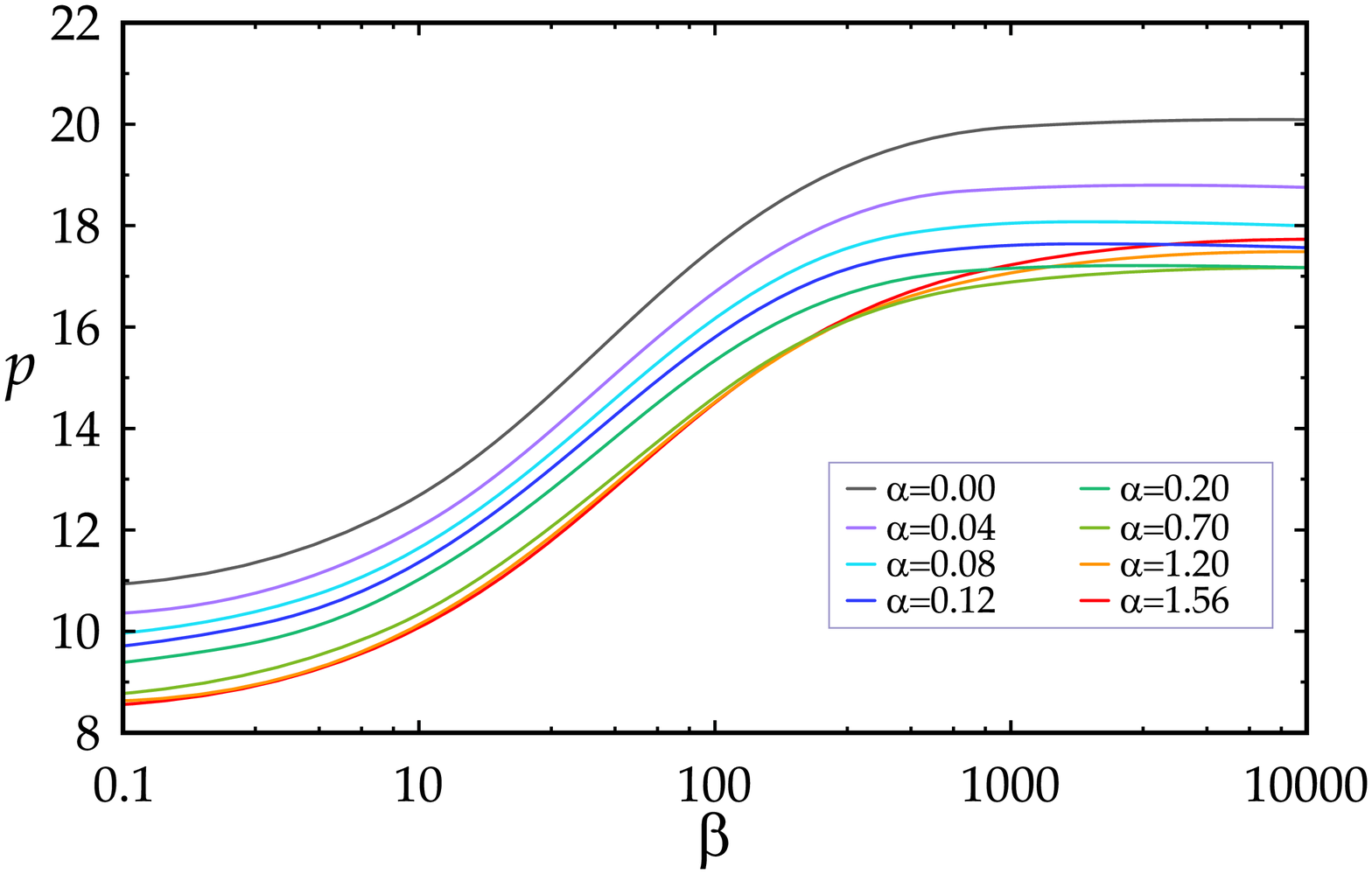}
    \caption{\footnotesize{
Each curve, computed with the use of the equation \rf{wu1}, shows the critical flutter load in the limit of vanishing dissipation  as a function of the damping ratio $\beta$ for the discretized model with $N=2$ and $\chi=1$ and corresponds to a different mass ratio $\alpha$ (reported in the legend). {Note that at} large mass ratios $0.7 \lesssim \alpha \le \pi/2$ the curves form a dense family.
}}
    \label{galerkin_betap}
  \end{center}
\end{figure}

Finally, {using equation \rf{wu1}, the critical flutter load in the limit of vanishing dissipation is plotted} in Fig.~\ref{galerkin_betap} as a function of the damping ratio $\beta$, for different mass ratios $\alpha \in [0, \pi/2]$.
{It is worth noting that} in the range  $0.7 \le \alpha \le \pi/2$ the curves form a dense family.
According to Fig.~\ref{figp3}(c), for $0.342716\le\alpha_0\le \pi/2$  the stabilizing damping ratio $\beta_0$ is negative and tends to infinity as $\alpha_0 \rightarrow  +0.342716\ldots$, which corresponds to the vertically oriented cusp in Fig.~\ref{figp2}(c).

\subsection{Agreement with the solution of the boundary eigenvalue problem \rf{beckeq}}\label{SectionPT11}

{When $N$ is increased}, the eigenvalues, eigenvectors, and stability boundary based on the finite-dimensional {approximation} \rf{eip} converge to {those solutions
of the eigenvalue problem \rf{beckeq}}.
However, already the $N=2$ approximation is in an excellent
{
qualitative agreement and in a very reasonable quantitative agreement with the solution of  \rf{beckeq}.
}
For completeness, {Appendix \ref{app_B} reports the perturbation formulas for the singular flutter boundary, which are} valid for arbitrary dimension $N$ of the discretized model.

                      \begin{figure}[bt!]
    \begin{center}
   \includegraphics[angle=0, width=\columnwidth]{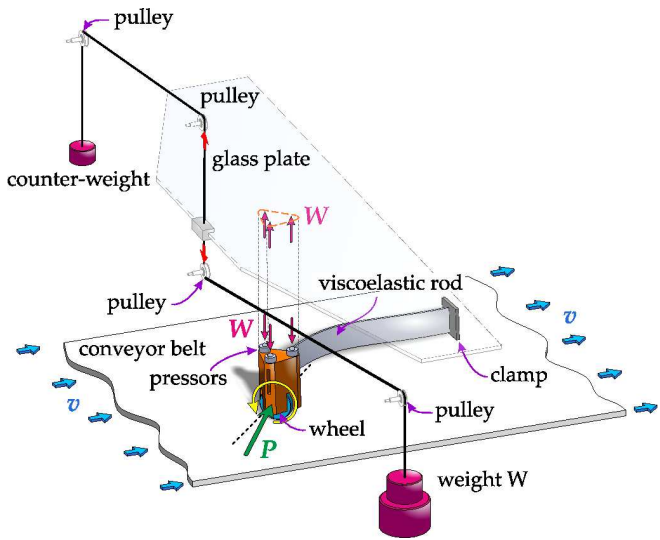}
    \end{center}
    \caption{\footnotesize{The sketch of the `flutter machine' producing the frictional partial
		follower load $P$ at the free end of the cantilevered viscoelastic Pfl\"uger column.}}
    \label{fig3}
    \end{figure}

\section{Experimental detection of the singular flutter limit}

\subsection{Experimental realization of the Pfl\"uger column}
Inspired by the Ziegler set-up \cite{BN2011}, a new mechanical device (Fig. \ref{fig3}) has been designed and realized to induce a follower force at the end of a Pfl\"uger column. The force (whose magnitude is continuously acquired with a miniaturized load cell) is produced by friction generated through sliding of a freely rotating wheel against a conveyor belt and can be calibrated as proportional (through the Coulomb friction rule) to {a} vertical load (provided via frictionless contact with a glass plate{, loaded through a pulley system}) pressing the wheel against the conveyor belt (which was running at a constant speed of $0.1$ m/s in all experiments) \cite{sm}.


\subsection{Identification of internal and external damping}\label{identification}

During vibration of a rod two types of dissipations arise: an external (due to the air drag) and an internal (due to the viscosity of the constitutive material of the rod) damping. Often external and internal damping are condensed in a single coefficient, but it was shown \cite{SW1975, T2016} that for problems of flutter a careful distinction has to be maintained between the different sources of damping, as both strongly influence results. Therefore, experiments were performed to identify the two damping parameters introduced in the model, namely, a viscous modulus $E^*$ (modelling the internal damping) and an air drag coefficient $K$ (corresponding to a distributed external damping). To this purpose, the viscoelastic rod used for the flutter experiments was mounted on a shaker in a cantilever configuration
{
and the acceleration of its free end measured when the basis was imposed a sinusoidal displacement of a frequency corresponding to the first two modes of resonance.
Results from these experiments were used with a modified logarithmic decrement approach detailed in Appendix \ref{app_C}, to obtain the following values of the internal and external
damping coefficients: $E^*=2.139796\cdot 10^6\,kg\, m^{-1} s^{-1}$ and $K=1.75239\cdot10^{-5}\,kg\, m^{-1} s^{-1}$.}


              \begin{figure*}[bt!]
    \begin{center}
    \includegraphics[angle=0, width=\textwidth]{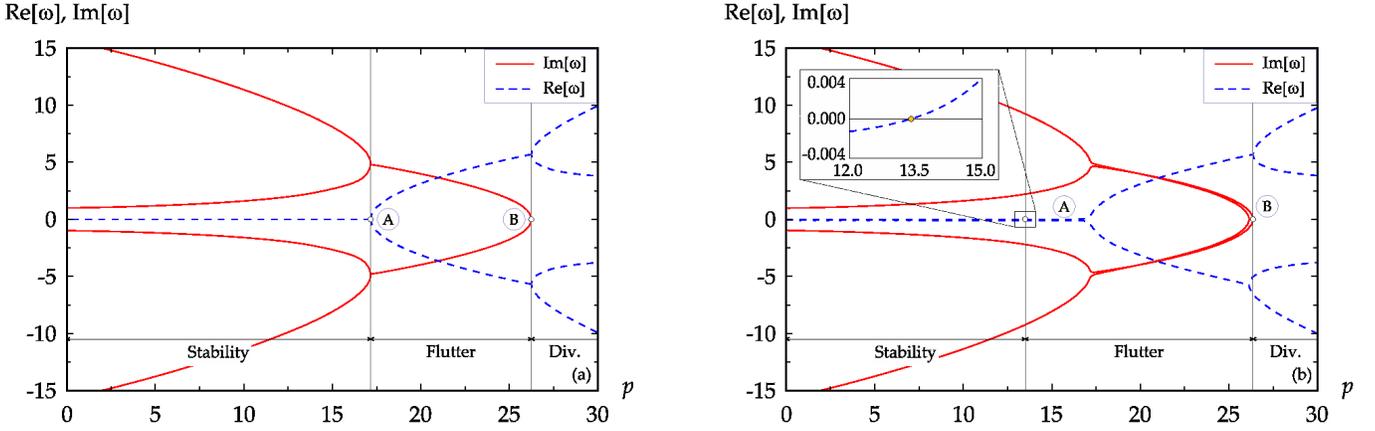}
    \end{center}
    \caption{Pulsation (red solid curves) and growth rates (blue dashed curves) for the Pfl\"uger column versus the dimensionless load $p$ (a) without damping and (b) in the presence of a Kelvin-Voigt damping for the material ($\eta$) and air drag ($\gamma$), demonstrating the drop in the onset of flutter. The plots were obtained with the parameters {representative} of sample 5 in Table~\ref{table_beam}. }
    \label{fig2}
    \end{figure*}

\subsection{Detection of the singular limit {for} the flutter onset}

Our experiments are compared with the numerical solution of the boundary eigenvalue problem \rf{beckeq}.
The roots of the characteristic equation \rf{trans_eq} are the  eigenvalues {
$\omega$ governing} the vibrations of the Pfl\"uger column. The first two eigenvalues with their conjugates are plotted in Fig.~\ref{fig2} versus the load $p$, with all the other parameters kept fixed. In the absence of both the Kelvin-Voigt damping ($\eta$) and the air drag ($\gamma$), the Pfl\"uger column is a reversible system and loses stability by flutter via collision of imaginary eigenvalues in a linear reversible-Hopf bifurcation, Fig.~\ref{fig2}(a). In the presence of the two dissipation mechanisms, the merging of modes is imperfect, thus yielding flutter through the classical Hopf bifurcation at a value of $p$  significantly lower than in the case when the dissipation source is absent, Fig.~\ref{fig2}(b).  The theory of the previous section predicts that when the damping coefficients tend to zero while their ratio is kept constant, a limiting value of the flutter onset is reached, which generically differs from the flutter onset of the undamped column, thus justifying the numerical results of \cite{T2016}.

The critical flutter load for the Pfl\"uger column was experimentally investigated covering a wide range of values of the mass ratio $\alpha$, Table~\ref{table_beam}.
Note that, since $E^*$ and $K$ are constant, the geometry of the tested rods parameterizes the dimensionless damping coefficients $\eta$ and $\gamma$ according to Eqs.~\rf{dimensionless}, so that  different values of $\gamma$ and $\eta$ are obtained for rods of different length $(l)$ and thickness $(b)$.

\begin{figure}[ht!]
    \begin{center}
    \includegraphics[angle=0, width=\columnwidth]{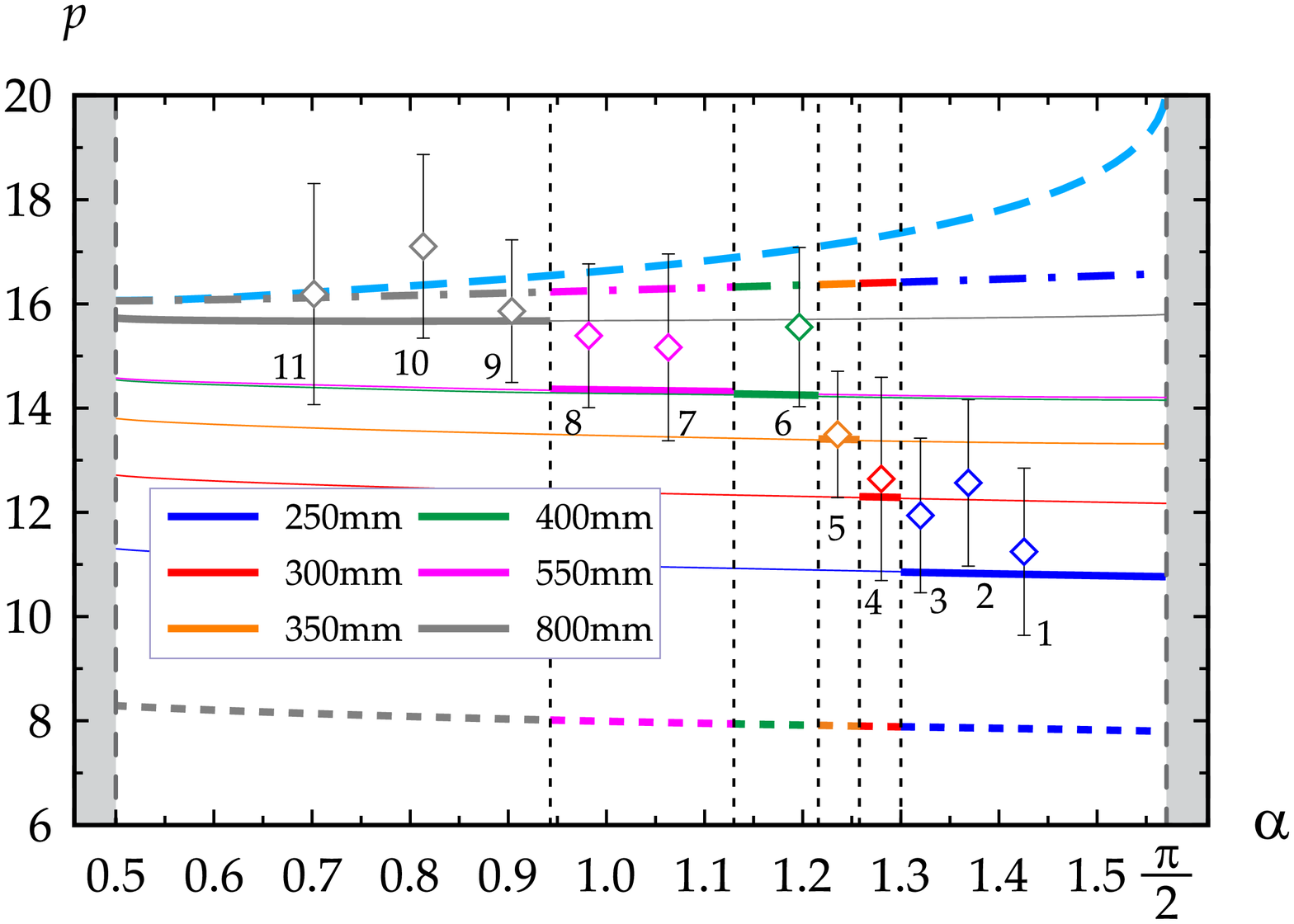}
    \end{center}
    \caption{Critical flutter load $p$ versus mass ratio $\alpha$. Theoretical predictions based on Eq.~\rf{trans_eq} are plotted (the upper dashed curve) when damping is absent, when only external ($\gamma$, dot-dashed lines) or internal ($\eta$, lower dashed lines) damping is present, and (solid lines) when both damping mechanisms are present. Experimental results are marked by diamonds with error bars. The tested samples are numerated and their characteristics reported in Table \ref{table_beam}.}
    \label{fig4}
    \end{figure}

The results of the measurements, together with the numerical calculations \cite{T2016,P1955}, are shown in Fig.~\ref{fig4} for eleven samples (see  Table~\ref{table_beam}) in the plane $p$ versus $\alpha$. Theoretical critical curves, pertaining to samples of different lengths and thicknesses, are plotted and highlighted for the relevant intervals of $\alpha$. These boundaries are well-separated from the flutter boundary of the undamped system, represented by the upper dashed curve. In cases when either $\eta=0$ (the dot-dashed curves) or $\gamma=0$ (the lower dashed curves) the difference between the flutter boundaries corresponding to samples of various geometry is
{hardly visible}, as it should be, in agreement with the theory, when the damping coefficients are very small \cite{T2016,P1955,B1956,KV2010,Z1952}. In contrast, when both damping mechanisms are taken into account, the critical curves dramatically differ for samples of different length and thickness. This is because the \textit{ratio} $\beta=\gamma/\eta=(K/E^*)(l^4/J)$ between the two damping coefficients increases almost 25 times from the first sample to the eleventh (see Table~\ref{table_beam}), although the damping coefficients $\gamma$ and $\eta$ vary weakly with the sample geometry.

\begin{table}
\centering
\begin{adjustbox}{width=0.46\textwidth}
\begin{tabular}{c|cccccccc}
\multirow{ 2}{*}{Rod} & $b$  & $l$  & $J$ & $M$  & $\alpha$ & $\eta,\times 10^{-3}$ & $\gamma,\times 10^{-3}$ & $\beta$ \\
 & [mm] & [mm]  & [mm$^4$] & [kg] & [-] & [-] & [-] & [-] \\
\hline
1    & 1.90  & 250 & 13.72 & 0.105 & 1.426 & 1.059 & 24.71   & 23.33    \\
2    & 1.90  & 250 & 13.72 & 0.075 & 1.369 & 1.059 & 24.71   & 23.33    \\
3    & 1.90  & 250 & 13.72 & 0.060 & 1.320 & 1.059 & 24.71   & 23.33    \\
4    & 1.90  & 300 & 13.72 & 0.060 & 1.280 & 0.746 & 36.06   & 48.33    \\
5    & 1.92  & 350 & 14.16 & 0.060 & 1.236 & 0.557 & 48.37   & 86.84    \\
6    & 1.95  & 400 & 14.83 & 0.060 & 1.196 & 0.439 & 62.13   & 141.5  \\
7    & 2.98  & 550 & 52.93 & 0.089 & 1.063 & 0.348 & 50.76   & 145.9  \\
8    & 2.98  & 550 & 52.93 & 0.075 & 0.982 & 0.348 & 50.76   & 145.9  \\
9    & 3.07  & 800 & 57.87 & 0.089 & 0.903 & 0.177 & 102.5 & 579.3  \\
10  & 3.07  & 800 & 57.87 & 0.075 & 0.813 & 0.177 & 102.5 & 579.3  \\
11  & 3.07  & 800 & 57.87 & 0.060 & 0.702 & 0.177 & 102.5 & 579.3  \\
\end{tabular}
\end{adjustbox}
\caption{Characterization of the different samples tested. Rods for all the 11 samples have identical height, $h=24$ mm.}\label{table_beam}
\end{table}

          \begin{figure}[ht!]
    \begin{center}
    \includegraphics[angle=0, width=\columnwidth]{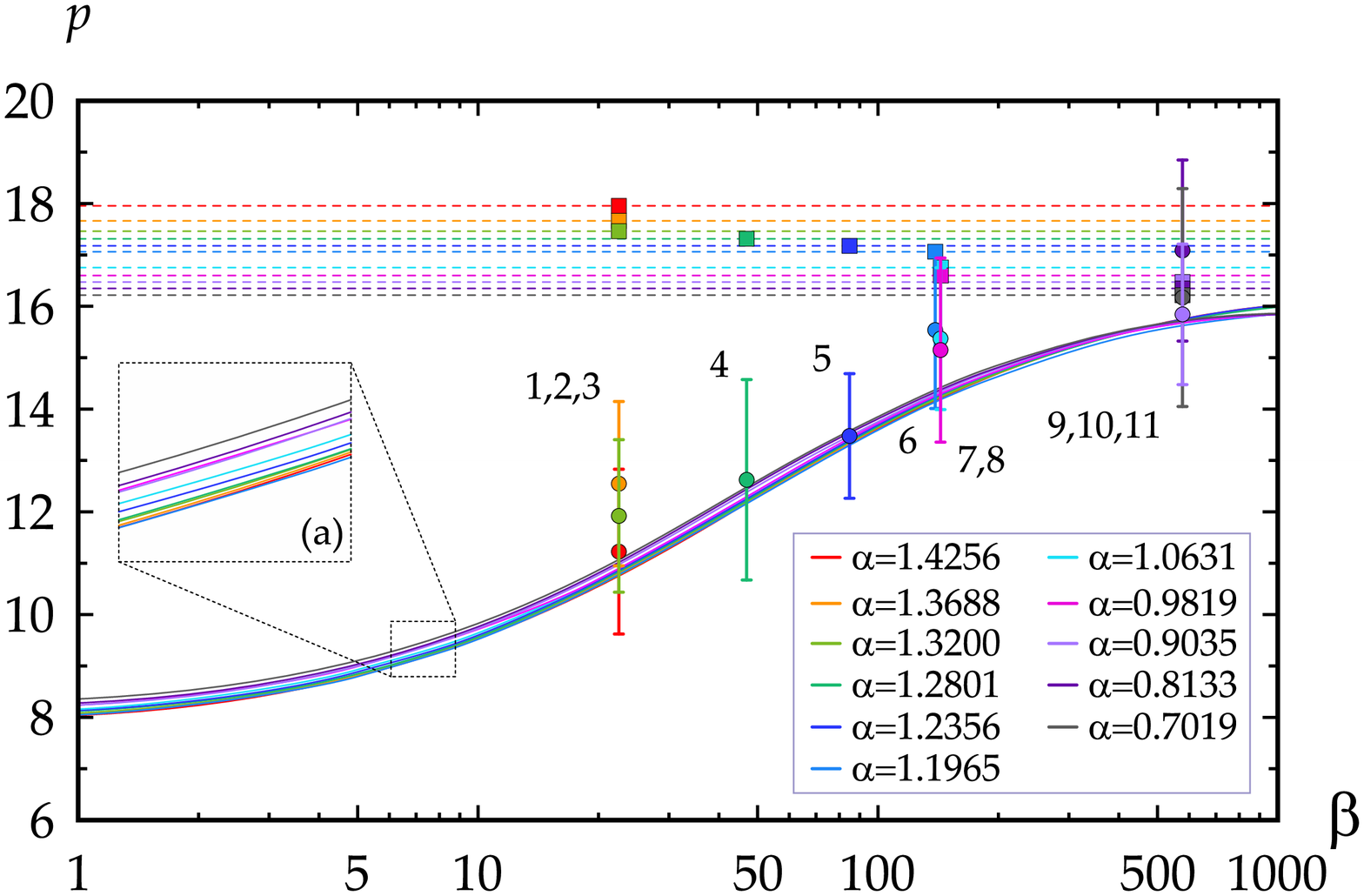}
    \end{center}
    \caption{Solid curves mark the critical flutter load versus damping ratio $\beta=\gamma/\eta$ at different values of mass ratio $\alpha$ and corresponding fixed values of $\eta$, see Table~\ref{table_beam}. The experimental data are shown by spots with error bars. Dashed lines indicate the critical flutter load of the undamped Pfl\"uger column for the same values of $\alpha$.}
    \label{fig5}
    \end{figure}

Assuming $\gamma=\beta \eta$ in equation \rf{trans_eq} and fixing $\eta$ to be one of the values reported in Table~\ref{table_beam},
the flutter boundary is plotted in Fig.~\ref{fig5} in the $p$ versus $\beta$ representation. Since for every length and thickness the critical flutter load depends weakly on $\alpha$, see Fig.~\ref{fig4}, the flutter boundaries in Fig.~\ref{fig5}, inset (a), are situated very close to each other (cf. Fig.~\ref{galerkin_betap}).
If the results of the measurements are superimposed, the experimental points  perfectly fit this family of boundaries, within the error bands. Both the theoretical curves and the experimental points lie below the critical values of the undamped system for all values of $\alpha$. Nevertheless, the critical flutter load of the weakly damped Pfl\"uger column is very sensitive to the damping ratio and increases as $\beta$ increases with the tendency to touch the lowest of the ideal flutter boundaries at $\beta > 1000$, where the critical loads of the damped and undamped system tend to coincide (within the error bands), Fig.~\ref{fig5}.

\subsection{The flutter modes}

\begin{figure*}[ht!]
  \begin{center}
     \includegraphics[width=\textwidth]{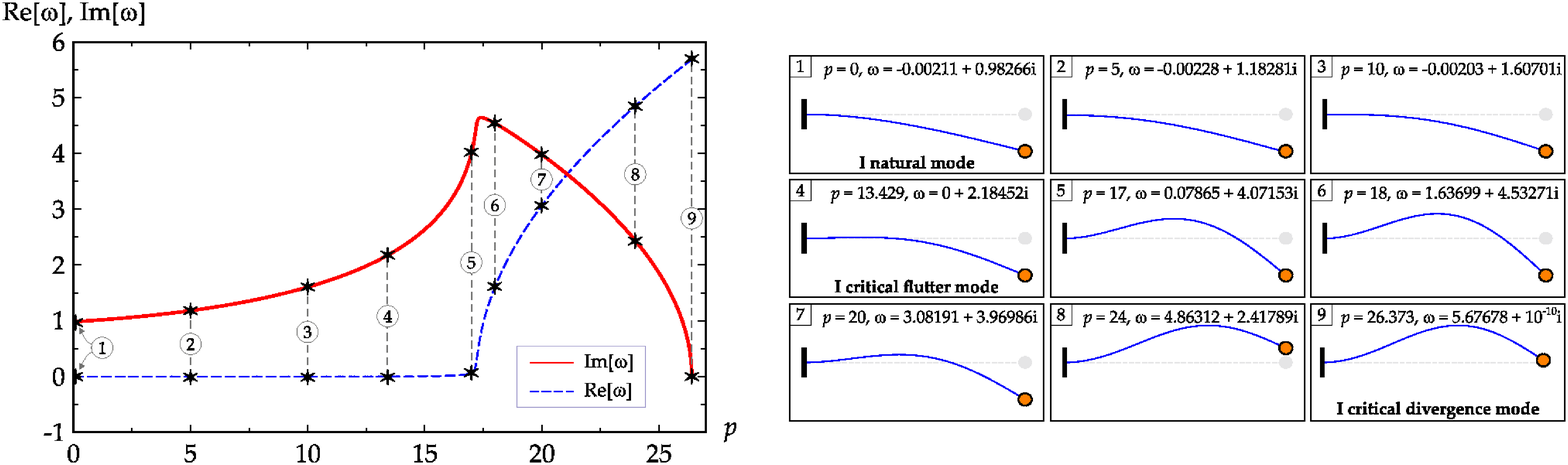}
    \caption{\footnotesize{
Real (blue dashed curve) and imaginary (red solid curve) part of the eigenfrequencies associated to the first (lower frequency) flutter branch. Each number corresponds to a value of the tangential load $p$ for which the relevant eigenvector is computed and reported on the right in separate boxes. The vibrations numbered 1 to 3 are stable. Flutter instability first occurs at the load for which the mode numbered 4 is reported.
}}
    \label{eigenvectors1}
  \end{center}
\end{figure*}

The analysis of the experiments is complemented by the determination of the flutter modes, which can be pursued by calculating the eigenvectors associated to the eigenvalues solutions of equation (7). The knowledge of the flutter modes is in fact useful to identify the shape of the vibrating rod during experiments.
The analysis of the eigenvectors is reported in Fig. \ref{eigenvectors1}, relative to the first (lower frequency) vibration branch for the sample n. 5 of Table I, with dimensionless dampings $\eta=0.557\cdot10^{-3}$ and $\gamma=48.368\cdot10^{-3}$.
All modes 1-3 in the figures refer to stable vibrations, while the onset of flutter corresponds to the mode numbered 4 and the onset of divergence to the mode numbered 9.

{
It is evident from Fig. \ref{eigenvectors1}(1) that the shape of the vibration mode corresponds (as it should be) at null $p$ to the free vibrations of a cantilever rod with a concentrated mass on its tip, vibrating at first resonance frequency.
When the load $p$ increases beyond the threshold of the classical-Hopf bifurcation and approaches the higher value of the load corresponding to the threshold of the
reversible-Hopf bifurcation in the undamped case, the vibrations become more and more similar to the second vibration mode of the free cantilever rod. This is not surprising in view of the fact that in the undamped case the eigenvectors of the first and the second mode merge at the flutter threshold because of the formation of a double imaginary eigenvalue with the Jordan block.}
In all the performed experiments the modes sketched in Fig. \ref{eigenvectors1} have been observed.

\section{Conclusion}
The theoretically predicted  singular limiting behavior {for the onset of} the classical Hopf bifurcation has been detected and can now be considered as experimentally confirmed for a nearly-reversible system in the limit of vanishing dissipation.

{This effect has been both theoretically and experimentally analyzed} on a classical paradigmatic model of a nearly-reversible system{, namely, }
the Pfl\"uger viscoelastic column moving in a resistive medium under the action of a tangential follower force. For the theoretical treatment the continuous non-self-adjoint boundary eigenvalue problem has been Galerkin-discretized and reduced to a finite-dimensional matrix eigenvalue problem. With the use of perturbation theory of multiple eigenvalues, explicit expressions for the critical flutter load with and without dissipation have been derived thus proving the Whitney umbrella singularity {at} the interface between the classical Hopf bifurcation of the dissipative Pfl\"uger system and the reversible-Hopf bifurcation of its undamped version. The conducted experiments with the laboratory realization of the Pfl\"uger column confirmed the high sensitivity of the flutter onset {to} the damping ratio and accurately fitted both the theoretically and numerically predicted laws.

The designed, manufactured and tested `flutter machine' opens a way to dedicated experiments on dissipation-induced instabilities with multiple damping mechanisms in a controlled laboratory environment.








\textit{Acknowledgments--} We thank I. Hoveijn for useful discussions. The authors gratefully acknowledge financial support from the ERC Advanced Grant ERC-2013-ADG-340561-INSTABILITIES.

\begin{appendix}
\section{Discretization}\label{App A}
\subsection{Adjoint boundary eigenvalue problems}

{The boundary eigenvalue problem for the Pfl\"uger column with partial follower load is given by equation (\ref{beckeq}).
The problem is self-adjoint only} for $\chi=0$ and non-self-adjoint otherwise. Indeed, {integration by parts of the differential equation \rf{beckeq} together with the boundary conditions lead to the following} adjoint boundary eigenvalue problem
\ba{aep}
&(1 +\eta \bar \omega) \tilde w'''' +p \tilde w''+(\gamma \bar \omega +\bar \omega^2)\tilde w=0,&\nn\\
&\tilde w(0)=\tilde w'(0)=0,\quad \tilde w''(1)(1+\eta \bar \omega)+\chi p\tilde w(1)=0\nn,& \\
&(1+\eta \bar \omega)\tilde w'''(1)+p\tilde w'(1)- \tilde w(1)\omega^2 \tan \alpha=0.&
\ea
The problem \rf{aep} coincides with  \rf{beckeq} only for $\chi=0$. Otherwise, the boundary conditions of the two problems differ.

\subsection{Variational principle}\label{SectionPT2}

{Let us consider now the} functional
\ba{fun}
&I(\tilde v, \tilde w)=\int_0^1\left[(1+\eta \omega)\tilde v''''\tilde w +p \tilde v'' \tilde w +(\gamma \omega +\omega^2)\tilde v \tilde w \right]d\xi.&\nn\\
\ea
Integrating by parts the first two terms in equation \rf{fun} and {accounting for} the boundary conditions for the problems \rf{beckeq} and \rf{aep},
{leads to}
\ba{bp}
&\int_0^1(\tilde v''')'\tilde w d\xi=\int_0^1\tilde v''\tilde w'' d\xi
+\tilde v'''(1)\tilde w(1),&\nn\\
&\int_0^1 (\tilde v')' \tilde w d\xi=-\int_0^1 \tilde v' \tilde w' d\xi+\tilde v'(1)\tilde w(1).&
\ea


On the other hand, {the last  of the boundary conditions \rf{beckeq} provides}
$$
(1+\eta \omega)\tilde v'''(1)+p\tilde v'(1)=\chi p\tilde v'(1)+ \tilde v(1)\omega^2 \tan \alpha.
$$
Hence,
\ba{pt1}
&I=\int_0^1\left[(1+\eta \omega)\tilde v''\tilde w '' -p \tilde v' \tilde w' +(\gamma \omega +\omega^2)\tilde v \tilde w \right]d\xi&\nn\\
&+\tilde v(1)\tilde w(1)\omega^2 \tan \alpha +\chi p \tilde v'(1)\tilde w(1).&
\ea
Stationarity of this functional with respect to arbitrary smooth variations
$\delta \tilde v$, $\delta \tilde w$, {which satisfy} kinematic boundary conditions, is equivalent to the boundary value
problems \rf{beckeq}, \rf{aep}.

\subsection{Discretization and reduced finite-dimensional model}\label{SectionPT3}

{Let us consider} solutions to the self-adjoint problems \rf{beckeq} and \rf{aep}, with $\chi=0$, $p=0$, $\eta=0$, $\gamma=0$, and $\alpha=0$
\begin{widetext}
\ba{basf}
\tilde v_j=\tilde w_j &=& \left|\frac{\sin\sqrt{\omega_j}}{1+(-1)^j\cos\sqrt{\omega_j}}\right| \left[\sin(\xi\sqrt{\omega_j})-\sinh(\xi\sqrt{\omega_j})-
\frac{\sin(\sqrt{\omega_j})+\sinh(\sqrt{\omega_j})}{\cos(\sqrt{\omega_j})+\cosh(\sqrt{\omega_j})}\left(\cos(\xi\sqrt{\omega_j})-
\cosh(\xi\sqrt{\omega_j})\right)\right],\nn\\
\ea
\end{widetext}
where $\omega_j$ is a root of the characteristic equation
$$
\cos(\sqrt{\omega})\cosh(\sqrt{\omega})+1=0,
$$
{which provides for} instance,
\ba{freq}
\omega_1&=&3.516015269, \quad \sqrt{\omega_1}=1.875104069\nn\\
\omega_2&=&22.03449156, \quad \sqrt{\omega_2}=4.694091132\nn\\
&\ldots&\nn\\
\omega_n&=&\frac{\pi^2}{4}(2n-1)^2,\quad \sqrt{\omega_n}=\frac{\pi}{2}(2n-1).
\ea
The functions \rf{basf} are orthogonal and normalized as follows:
$$
\int_0^1 \tilde v_i(\xi) \tilde v_j(\xi) d\xi =0,\quad i\ne j; \quad \int_0^1 \tilde v_i(\xi) \tilde v_i(\xi) d\xi =1.
$$
Therefore, {the eigenmodes $\tilde v$ and $\tilde w$ can be represented} in the form of the expansions
\be{approx}
\tilde v\approx \sum_{j=1}^N a_j \tilde v_j(\xi), \quad \tilde w\approx \sum_{j=1}^N b_j \tilde w_j(\xi),
\ee
where $\tilde w_j=\tilde v_j$.

Substituting the expansions \rf{approx} into the functional \rf{pt1} yields the discretized  version of the functional \rf{pt1}
\ba{ptd}
I_N&=&\omega^2\sum_{i=1}^N \sum_{j=1}^N a_i  b_j  \left(\int_0^1 \tilde v_i \tilde v_j d\xi+\tilde v_i(1)\tilde v_j(1) \tan \alpha\right)\nn\\
&+&\omega\sum_{i=1}^N \sum_{j=1}^N a_i  b_j  \int_0^1\left[\eta   \tilde v_i'' \tilde v_j ''  +\gamma  \tilde v_i \tilde v_j \right]d\xi\\
&+&\sum_{i=1}^N \sum_{j=1}^N a_i  b_j\left(\int_0^1 [\tilde v_i'' \tilde v_j ''-p   \tilde v_i'  \tilde v_j']d\xi +\chi p \tilde v_i'(1)\tilde v_j(1)\right).\nn
\ea

{The gradient} of the discretized functional, $I_N$, {calculated} with respect to the vector of coefficients ${\bf b}=(b_1,b_2,\ldots,b_N)$, and {equated to zero, provides} the discretized eigenvalue problem for the Pfl\"uger column
\be{dip1}
({\bf M}\omega^2+(\gamma{\bf D}_e+\eta{\bf D}_i)\omega+{\bf K}_1-p{\bf K}_2+\chi p{\bf N}){\bf a}=0,
\ee
where ${\bf a}=(a_1,a_2,\ldots,a_N)$ and the elements of the matrices are
\ba{mat}
&M_{ij}=\int_0^1 \tilde v_i \tilde v_j d\xi+\tilde v_i(1)\tilde v_j(1) \tan \alpha &\nn\\
&=\delta_{ij}+4(-1)^{i+j} \tan \alpha,&\nn\\
&D_{e,ij}=\int_0^1\tilde v_i \tilde v_j d\xi=\delta_{ij},
\quad D_{i,ij}=\int_0^1   \tilde v_i'' \tilde v_j '' d\xi=\delta_{ij}\omega_j^2,&\nn\\
&K_{1,ij}=\int_0^1   \tilde v_i'' \tilde v_j '' d\xi=\delta_{ij}\omega_j^2,\quad K_{2,ij}=\int_0^1   \tilde v_i' \tilde v_j ' d\xi,&\nn\\ &N_{ij}=\tilde v_i'(1)\tilde v_j(1)=\frac{4(-1)^{j+1}\sqrt{\omega_i}\sin\sqrt{\omega_i}}{1+(-1)^i\cos\sqrt{\omega_i}},&
\ea
with $\delta_{ij}$ denoting the Kronecker symbol. The entries of the matrix ${\bf K}_2$ in the explicit form are
\ba{k22}
&i\ne j:~~ K_{2,ij} = A \left(
\frac{\sqrt{\omega_j}\sin(\sqrt{\omega_i})}{\cos(\sqrt{\omega_i})(-1)^i+1}-\frac{\sqrt{\omega_i}\sin(\sqrt{\omega_j})}{\cos(\sqrt{\omega_j})(-1)^j+1}\right),
&\nn\\
&i=j:~~ K_{2,jj}=\frac{\omega_j\left((-1)^j-\cos\sqrt{\omega_j}\right)-2\sqrt{\omega_j}\sin{\sqrt{\omega_j}}}{\cos\sqrt{\omega_j}+(-1)^j},&\nn\\
\ea
where $A=\frac{4\sqrt{\omega_i\omega_j}}{(-1)^i\omega_i-(-1)^j\omega_j}$.
All the matrices are real. {In addition}, the matrices of mass, $\bf M$, external damping, ${\bf D}_e$, internal damping, ${\bf D}_i$, and stiffness, ${\bf K}_1$ and ${\bf K}_2$, are symmetric. The matrix of nonconservative positional forces with non-zero curl, $\bf N$, is real and non-symmetric. Note that $\det{\bf M}=1+4N \tan \alpha >0$.

\section{Perturbation formulas for arbitrary $N$}\label{app_B}

The eigenvalue problem \rf{eip2} can be formulated as the eigenvalue problem
$$
{\bf L}(\omega,{\bf k}){\bf a}=0
$$
for the matrix polynomial
$$
{\bf L}(\omega,{\bf k}):={\bf A}(p,\chi)+{\bf D}(\gamma,\eta)\omega+{\bf M}(\alpha)\omega^2,
$$
where ${\bf k}=(p,\chi,\gamma,\eta,\alpha)$ is a vector of parameters. {The adjoint matrix polynomial ${\mathbf L}^{\dagger}={\mathbf A}^T+{\bf D}{\overline \omega}+{\mathbf M}{\overline \omega}^2$ is introduced}, so that
 $({\mathbf L}{\mathbf a},{\mathbf b})=({\mathbf a},{\mathbf L}^{\dagger} {\mathbf b})$, where the inner product is defined as $({\bf a},{\bf b})=\overline{\bf b}^T{\bf a}$.
{With this definition, the adjoint eigenvalue problem can be rewritten as}
$$
{\mathbf L}^{\dagger}(\overline{\omega},{\bf k}) {\mathbf b}=0.
$$




{Let us assume that, for} the values of the parameters $\chi=\chi_0$, $\alpha=\alpha_0$, $\gamma=0$, $\eta=0$, and $p=p_0$, {an algebraically double imaginary eigenvalue $\omega_0=i \sigma_0$ exists} with the Jordan block {which} satisfies the following equations
\ba{Jordan}
{\mathbf A}_0{\mathbf a}_0-\sigma_0^2 {\mathbf M}_0{\mathbf a}_0&=&0\nn, \\
{\mathbf A}_0{\mathbf a}_1-\sigma_0^2 {\mathbf M}_0{\mathbf a}_1&=&-2i\sigma_0 {\mathbf M}_0{\mathbf a}_0,
\ea
where ${\mathbf a}_0$ is an eigenvector and ${\mathbf a}_1$ is an associated vector at $\omega_0$.
Then, an eigenfunction ${\mathbf b}_0$ and an associated function ${\mathbf b}_1$ at the complex-conjugate eigenvalue $\overline{\omega}_0=-i\sigma_0$ are governed by the adjoint equations
\ba{Jordan1}
{\mathbf A}_0^T{\mathbf b}_0-\sigma_0^2 {\mathbf M}_0{\mathbf b}_0&=&0\nn, \\
{\mathbf A}_0^T{\mathbf b}_1-\sigma_0^2 {\mathbf M}_0{\mathbf b}_1&=&2i\sigma_0 {\mathbf M}_0{\mathbf b}_0.
\ea
Note the orthogonality {between} the eigenvectors, that is
\be{ort}
({\bf M}_0{\bf a}_0,{\bf b}_0)=0.
\ee

{When} the parameter $p$ is perturbed in the vicinity of $p_0$ as $p=p_0+\Delta p$, {an approach similar to that used for $N=2$ yields}
\ba{bifEig}
&\omega(p)=i\sigma_0\pm\sqrt{\Delta p\,\frac{i({\mathbf A}'_p{\mathbf a}_0,{\mathbf b}_0)}{2\sigma_0({\mathbf M}_0{\mathbf a}_1,{\mathbf b}_0)}}+o(\sqrt{|\Delta p|}),&\nn\\
&{\mathbf a}(p)={\mathbf a}_0\pm{\mathbf a}_1\sqrt{\Delta p\,\frac{i({\mathbf A}'_p{\mathbf a}_0,{\mathbf b}_0)}{2\sigma_0({\mathbf M}_0{\mathbf a}_1,{\mathbf b}_0)}}+o(\sqrt{|\Delta p|}),&\nn\\
&{\mathbf b}(p)={\mathbf b}_0\pm{\mathbf b}_1\sqrt{\Delta p\,\frac{i({\mathbf A}'_p{\mathbf a}_0,{\mathbf b}_0)}{2\sigma_0({\mathbf M}_0{\mathbf a}_1,{\mathbf b}_0)}}+o(\sqrt{|\Delta p|}),&
\ea
where ${\mathbf A}'_p=\left.\frac{\partial {\mathbf A}}{\partial p}\right|_{p=p_0}$.
Therefore,
the eigenvalues and eigenvectors of the undamped reversible system {can be approximated} in the vicinity of $p=p_0$, i.e. in the vicinity of the flutter boundary corresponding to the reversible-Hopf bifurcation.

Assume that at $p<p_0$ the eigenvalues of the undamped reversible system are imaginary, $\omega(p)=i\sigma(p)$, with an eigenvector
${\mathbf a}(p)$ and the eigenvector of the adjoint problem ${\mathbf b}(p)$. Then, at $p>p_0$ the eigenvalues \rf{bifEig} are complex-conjugate ({denoting} instability).
A dissipative perturbation with the matrix ${\mathbf D}(\eta,\gamma)$ where ${\mathbf D}(0,0)=0$ changes the eigenvalue $\omega(p)=i\sigma(p)$ as follows
\ba{lambdadis}
&\omega(p,\eta,\gamma)=\omega(p)&\nn\\
&-\frac{({\mathbf D}'_{\eta} {\mathbf a}(p), {\mathbf b}(p))\eta+({\mathbf D}'_{\gamma} {\mathbf a}(p), {\mathbf b}(p)) \gamma}{2({\mathbf M}_0{\mathbf a}(p), {\mathbf b}(p))}+o(|\eta|,|\gamma|).&
\ea
{The following condition for the imaginary eigenvalue is assumed to hold}
\be{ay}
({\mathbf D}'_{\eta} {\mathbf a}(p), {\mathbf b}(p))\eta+({\mathbf D}'_{\gamma} {\mathbf a}(p), {\mathbf b}(p)) \gamma=0 ,
\ee
{so that the eigenvalue remains imaginary after a dissipative perturbation}. This means that the neutral stability surface is not abandoned after the dissipative perturbation. Using the perturbation formulas \rf{bifEig} for ${\mathbf a}(p)$ and ${\mathbf b}(p)$ in \rf{ay}, introducing the damping ratio $\beta=\gamma/\eta$, and defining
\be{beta0n}
\beta_0=-\frac{({\mathbf D}'_{\eta} {\mathbf a}_0, {\mathbf b}_0)}{({\mathbf D}'_{\gamma} {\mathbf a}_0, {\mathbf b}_0)}=-\frac{({\mathbf D}_i {\mathbf a}_0, {\mathbf b}_0)}{({\mathbf a}_0, {\mathbf b}_0)} ,
\ee
{the following quadratic approximation in $\beta$ can be found} to the critical flutter load in the limit of vanishing dissipation
\begin{widetext}
\be{wun}
p=p_0+\frac{2\sigma_0({\mathbf M}_0{\mathbf a}_1,{\mathbf b}_0)}{i({\mathbf A}'_p{\mathbf a}_0,{\mathbf b}_0)}\left(\frac{({\mathbf D}'_{\gamma}{\mathbf a}_0,{\mathbf b}_0)}{[({\mathbf D}'_{\gamma}{\mathbf a}_0,{\mathbf b}_1)+({\mathbf D}'_{\gamma}{\mathbf a}_1,{\mathbf b}_0)]\beta_0+[({\mathbf D}'_{\eta}{\mathbf a}_0,{\mathbf b}_1)+({\mathbf D}'_{\eta}{\mathbf a}_1,{\mathbf b}_0)]} \right)^2(\beta-\beta_0)^2.
\ee
\end{widetext}

From the orthogonality of eigenvectors \rf{ort} and the expression for the mass matrix ${\bf M}_0={\bf I}+4{\bf M}_1\tan \alpha_0$  it follows immediately that the denominator in \rf{beta0n} vanishes at $\alpha_0=0$, thus confirming that in the case of the Beck column the external air drag damping is stabilizing. Now {this result
has been established} for the discretized model of the Pfl\"uger column of arbitrary dimension $N$.


In the case of $N=2$, $\chi_0=1$, $\alpha_0=0.1$, $p_0\approx17.83368$, $\sigma_0\approx 9.366049$, {the following vectors are obtained}
\ba{vectors}
&{\bf a}_0\approx \left(
             \begin{array}{c}
               0.720378 \\
               1 \\
             \end{array}
           \right),\quad
{\bf a}_1\approx -i\left(
             \begin{array}{c}
               0.225316 \\
               0.478780 \\
             \end{array}
           \right),&\nn\\
&{\bf b}_0\approx \left(
             \begin{array}{c}
               -1.828847 \\
               1 \\
             \end{array}
           \right),~~
{\bf b}_1\approx i\left(
             \begin{array}{c}
               -0.3423417 \\
               0.505899 \\
             \end{array}
           \right).&
\ea

With these vectors the formula \rf{bifEig} exactly reproduces equation \rf{traf1}. The formula \rf{beta0n} provides $\beta_0\approx1478.074$ in full accordance with equation \rf{beta0} in the case of $N=2$. Finally, equation \rf{wun} exactly reproduces equation \rf{apwu}.

For $N>2$ the procedure is the same: one only needs to find the vectors ${\bf a}_0$, ${\bf a}_1$, ${\bf b}_0$, ${\bf b}_1$ solving \rf{Jordan} and \rf{Jordan1} with the corresponding $N\times N$ matrices which entries are given by equations \rf{mat} and \rf{k22}.

\section{Modified logarithmic decrement approach}\label{app_C}

\subsection*{Equations of motion}\label{Section1}

A viscoelastic rod is considered, made up of a  material which follows the Kelvin-Voigt model
\be{kelvinvoigt}
\sigma_z = E \varepsilon_z + E^* \dot{\varepsilon}_z ,
\ee
where $\sigma_z$ and $\varepsilon_z$ are the longitudinal stress and strain, respectively, and $E$ and $E^*$ are the elastic and the viscous moduli.
In an Euler rod the strain is defined as
\be{strain}
\varepsilon_z = \frac{d\phi}{dz} y = \phi' y,
\ee
where $\phi'$ is the curvature and $y$ the coordinate orthogonal to the rod's {axis $x$}, so that the bending moment can be computed as
\ba{}
&\mathcal{M}=\int_A \sigma_z y dA = E \phi' \int_A y^2 dA+E^* \dot{\phi}' \int_A y^2 dA&\nn\\
 &= EJ \phi'+E^* J \dot{\phi}',&
\ea
and rewritten in terms of displacement $v(x,t)$ as
\be{eubern}
\mathcal{M}=-EJ v''-E^* J \dot{v}'' .
\ee

The equation governing the dynamics of a straight rod is
\be{din}
\mathcal{M}''=-p+m \ddot{v},
\ee
where $m$ is the mass density per unit length of the rod and $p$ is the transversal load per unit length, which can be identified with the sum of an applied load $f(t)$ and a force proportional (through a coefficient $K$) to the velocity $\dot{v}$, to model external damping. A substitution of Eq. (\ref{eubern}) into Eq. (\ref{din}) yields
\be{eq_gen}
EJ v^{IV}+E^* J \dot{v}^{IV}+K \dot{v}+m \ddot{v}= f(t).
\ee

A  sinusoidal excitation at the clamped end of a rod in a cantilever configuration can be modeled with a specific form of external load, namely
\be{f_t}
f(t)=m U_0 \bar{\omega}^2 \sin \bar{\omega} t ,
\ee
where $U_0$ is the amplitude of the displacement imposed at the clamp, which varies in sinusoidally in time with pulsation $\bar{\omega}$.


\subsection{Free vibration of a cantilever rod}\label{Section2}

The solution of Eq. (\ref{eq_gen}) with an imposed sinusoidal displacement in terms of $v(x,t)$ can be found exploiting the separation of variables
\be{sol_inf}
v\left(x,t\right)=\sum\limits_{n=1}^\infty Y_n(x)\cdot y_n(t) ,
\ee
where the function $Y_i(x)$ and $y_i(t)$ are mode functions, respectively, in space $x$ and in time $t$.
The force $f(t)$ acting on the rod plays a role only in the definition of the $y_i(t)$ modes.
Assuming a function of time $y(t)=\exp(-i \omega t)$ yields the characteristic equation
\ba{damped_Y}
&\sum\limits_{n=1}^\infty\left( 1-i\frac{\omega_n E^*}{E}\right)Y_n^{IV}-\left(\frac{m \omega_n^2}{EJ}+i \frac{\omega_n K}{EJ}\right)Y_n=0,&\nn\\
&\rightarrow \, \sum\limits_{n=1}^\infty Y_n^{IV}-\Lambda_n^4 Y_n=0,&
\ea
where $\Lambda_n^4$ is a real quantity (dimensionally equal to [length]$^{-4}$)
\be{}
\Lambda_n^4=\frac{m \omega_n^2+i \omega_n K}{EJ-i \omega_n E^* J}.
\ee

The solution to Eq. (\ref{damped_Y}) is a sum of periodic and hyperbolic functions
\ba{free_coef}
&Y(x)=\sum\limits_{n=1}^\infty Y_n(x)=\sum\limits_{n=1}^\infty C_{1,n}\sin \Lambda_n x + C_{2,n}\cos \Lambda_n x&\nn\\
&+ \sum\limits_{n=1}^\infty C_{3,n}\sinh \Lambda_n x + C_{4,n}\cosh \Lambda_n x& ,
\ea
where the constants $C_{i,n}$ depend on the boundary conditions. For a cantilever rod, the boundary conditions are
\be{bcbeck}
Y(0)=Y'(0)=Y''(l)=Y'''(l)=0.
\ee

A substitution of the boundary conditions  in Eq. (\ref{free_coef}) yields in a matrix form
\begin{widetext}
\be{sistema_1bc}
 \left[ \begin{array}{cccc}
0 & 1 & 0 & 1\\
\Lambda_n & 0 & \Lambda_n & 0\\
-\Lambda_n^2 \sin \Lambda_n l & -\Lambda_n^2 \cos \Lambda_n l & \Lambda_n^2 \sinh \Lambda_n l & \Lambda_n^2 \cosh \Lambda_n l\\
-\Lambda_n^3 \cos \Lambda_n l & \Lambda_n^3 \sin \Lambda_n l & \Lambda_n^3 \cosh \Lambda_n l & \Lambda_n^3 \sinh \Lambda_n l\\
\end{array} \right]
\left(
\begin{array}{c}
C_{1,n}\\
C_{2,n}\\
C_{3,n}\\
C_{4,n}\\
\end{array} \right)
=
\left(
\begin{array}{c}
0\\
0\\
0\\
0\\
\end{array} \right)
.
\ee
\end{widetext}

The first two equations yield
$$C_{4,n}=-C_{2,n},\quad C_{3,n}=-C_{1,n},$$ so that Eq. (\ref{sistema_1bc}) reduces to
\ba{sistema_1bc_red}
& (\sin \Lambda_n l +\sinh \Lambda_n l)C_{1,n}+  (\cos \Lambda_n l + \cosh \Lambda_n l)C_{2,n}=0&,\nn\\
& (\cos \Lambda_n l +\cosh \Lambda_n l)C_{1,n}-  (\sin \Lambda_n l - \sinh \Lambda_n l)C_{2,n}=0&\nn\\ .
\ea
Imposing the determinant of the matrix of the system \rf{sistema_1bc_red} to vanish provides
\be{trasc}
\cos \Lambda_n l \cosh \Lambda_n l=-1.
\ee
Eq. (\ref{trasc}) defines the $\Lambda_n$ values as
\ba{}
&\Lambda_1 l=1.875...,~~
\Lambda_2 l=4.694...,&\nn\\
&\Lambda_3 l=7.855...,~~\ldots,~~
\Lambda_n l=\frac{\pi}{2}(2n-1).\nn&
\ea

Now, the solution for the functions $Y_n(x)$ can be expressed in terms of one arbitrary constant $C_{1,n}$, so that
\ba{}
&C_{2,n}=-\frac{\sin \Lambda_n l+\sinh \Lambda_n l}{\cos \Lambda_n l + \cosh \Lambda_n l} C_{1,n}= \frac{\cos \Lambda_n l+\cosh \Lambda_n l}{\sin \Lambda_n l - \sinh \Lambda_n l} C_{1,n}& ,\nn
\ea
which leads to the general solution for the free vibrations of a cantilever rod expressed as an infinite sum of the following mode functions
\begin{widetext}
\be{free_undamped}
\begin{array}{lll}
Y_n(x)&=& C_{1,n} \left[ \sin \Lambda_n x -\sinh \Lambda_n x -\frac{\sin \Lambda_n l+\sinh \Lambda_n l}{\cos \Lambda_n l + \cosh \Lambda_n l} \left( \cos \Lambda_n x-\cosh \Lambda_n x\right) \right] \\ [5 mm]
&=& C_{1,n} \left[ \sin \Lambda_n x -\sinh \Lambda_n x +\frac{\cos \Lambda_n l+\cosh \Lambda_n l}{\sin \Lambda_n l - \sinh \Lambda_n l} \left( \cos \Lambda_n x-\cosh \Lambda_n x\right) \right].
\end{array}
\ee
\end{widetext}


\subsection{Properties of the function $Y_n(x)$}\label{Section3}

The free vibration shape equations $Y_n(x)$ satisfy the orthogonality relations
\be{}
\begin{array}{lcr}
\int_0^l Y_n(x)Y_k(x)dx=0 && \text{for}\,\, k\neq n.
\end{array}
\ee
Morover, equation (\ref{damped_Y}) {allows to write}
\begin{equation}
\label{farlocca}
Y^{IV}_n(x)=\Lambda_n^4 Y_n(x).
\end{equation}
It is expedient now to define the quantity
\begin{equation}
\Gamma_n =\int_0^l Y^2_n(x)dx,
\end{equation}
{so that equation (\ref{farlocca}) yields}
\begin{equation}
\Gamma_n \Lambda_n^4 =\int_0^l Y^{IV}_n(x)Y_n(x)dx.
\end{equation}


\subsection{Expression of $y(t)$ for a cantilever rod with a base motion excitation}\label{Section4}

The differential equations governing the sinusoidal motion of the clamped rod subject to the force $f(t)$, Eq. (\ref{f_t}), are
\ba{general}
 &\sum\limits_{n=1}^\infty Y_n^{IV}(x)y_n(t)+\frac{E^*}{E}Y_n^{IV}(x)\dot{y}_n(t)
+\frac{K}{EJ}Y_n(x)\dot{y}_n(t)&\nn\\&+\frac{m}{EJ}Y_n(x)\ddot{y}_n(t)=\frac{f(t)}{EJ}.&
\ea

In order to exploit the orthogonality property of the shape functions $Y_n(x)$, each term of the previous equation is multiplied by $Y_k(x)$ and integrated over the length of the rod $l$, which provides the expression
\ba{general_simp}
&\Gamma_n \Lambda_n^4 y_n(t)+\Gamma_n \left( \frac{K}{EJ}+\frac{E^*}{E}\Lambda_n^4\right)\dot{y}_n(t)&\nn\\
&+\Gamma_n \frac{m}{EJ}\ddot{y}_n(t)=F_n \frac{f(t)}{EJ},&
\ea
where $F_n=\int_0^l Y_n(x)dx$.
Eq. (\ref{general_simp}) reminds the equation of motion which governs a single-degree-of-freedom system with a mass $m_n$, a damper with constant  $c_n$ and a spring with stiffness $k_n$
\be{sdof}
m_n \ddot{y}_n(t)+c_n \dot{y}_n(t)+k_n y_n(t)=p_n \sin \bar{\omega} t,
\ee
where
\ba{}
&m_n=\Gamma_n\frac{m}{EJ},\quad c_n=\Gamma_n\left(\frac{K}{EJ}+\frac{E^*}{E}\Lambda_n^4 \right),&\nn\\
&k_n=\Gamma_n \Lambda_n^4,\quad p_n=F_n \frac{\rho U_0 \bar{\omega}^2}{EJ}.&
\ea
Another form of Eq. (\ref{sdof}) is
\be{sdof_simp}
\ddot{y}_n(t)+2\alpha_n \zeta_n \dot{y}_n(t)+\alpha_n^2 y_n(t)=a_n \sin \bar{\omega}t,
\ee
where
\ba{sdof_coeff}
&\alpha_n^2=\frac{k_n}{m_n}=\frac{EJ}{m}\Lambda_n^4, \quad 2\alpha_n \zeta_n=\frac{c_n}{m_n}=\frac{K}{m}+\frac{E^* J}{m}\Lambda_n^4,&\nn\\
&a_n=\frac{p_n}{m_n}= \frac{F_n}{\Gamma_n} U_0 \bar{\omega}^2.&
\ea

The solution of the differential equation (\ref{sdof_simp}) is expressed as the sum of the solution of the associated homogeneous equation and of a particular integral. The latter can be found in the form
\be{}
y_{n,part}(t)=A_n \sin \bar{\omega}t+B_n\cos \bar{\omega}t ,
\ee
where the coefficients $A_n$ and $B_n$ satisfy Eq. (\ref{sdof_simp}) and assume the form
\be{a_b_part}
\begin{array}{cc}
A_n=a_n \left[1-\left(\frac{\bar{\omega}}{\alpha_n} \right)^2 \right] N_n ,& B_n=-2 a_n \zeta_n  \left(\frac{\bar{\omega}}{\alpha_n} \right) N_n ,
\end{array}
\ee
in which $N_n$ is the {so-called \lq dynamic amplification factor'}
\be{daf}
N_n(\alpha_n,\zeta_n)=\frac{1}{\left[1-\left(\frac{\bar{\omega}}{\alpha_n} \right)^2 \right]^2+\left[2\zeta_n \frac{\bar{\omega}}{\alpha_n} \right]^2}.
\ee

The solution of the homogeneous equation is
\be{}
y_{n,hom}(t)=\exp(-\zeta_n \alpha_n t)\left( C_n \sin \alpha_{n,d}t+D_n \cos \alpha_{n,d}t\right) ,
\ee
where $\alpha_{n,d}=\alpha_n \sqrt{1-\zeta_n^2}$ are the damped pulsations of the system.\\
The coefficients $C_n$ and $D_n$ can be found by imposing the initial conditions
\be{}
\begin{array}{cc}
y_{n,tot}(0)=X_0, & \dot{y}_{n,tot}(0)=V_0 ,
\end{array}
\ee
in the complete solution of
\be{}
y_{n,tot}(t)=y_{n,hom}(t)+y_{n,part}(t),
\ee
which leads to the  expressions

\ba{}
&C_n=\frac{1}{\alpha_{n,d}}\left[X_0 \alpha_n \zeta_n+V_0+a_n \bar{\omega} N_n\left(\frac{\bar{\omega}^2}{\alpha_n^2}+2\zeta_n^2-1 \right) \right],&\nn\\
&D_n=X_0+2 a_n \zeta_n\frac{\bar{\omega}}{\alpha_n}N_n.&
\ea

\subsection{Relation between $\zeta_n$, $E^*$, $K$ }\label{Section5}

The relation between the damping ratio $\zeta_n$, the internal ($E^*$), and the external ($K$) damping is described by Eq. (\ref{sdof_coeff})$_2$, which can be rewritten as
\be{}
\zeta_n=\frac{1}{2\Lambda_n^2} \left(\frac{K}{J}+E^* \Lambda_n^4 \right)\sqrt{\frac{J}{m E}} .
\ee

 The problem of the identification of the two damping coefficients thus reduces to the quantification of the damping ratio $\zeta_n$ relative to two different modes.
 The logarithmic decay over $k$ cycles can be written as
 \be{log_decay}
 \zeta_n=\frac{\delta_k}{2\pi k \alpha_n/\alpha_{n,d}}\approx \frac{\delta_k}{2\pi k} ,
 \ee
where $\delta_k=\log\left( y_1/y_{k+1}\right)$.

The dimensionless internal and external damping coefficients can be finally expressed through the relations
\be{}
\begin{array}{cc}
\gamma=\frac{K l^2}{\sqrt{m E J}}, & \eta=\frac{E^* l^2}{\sqrt{m E J}}\frac{J}{l^4}.
\end{array}
\ee

\end{appendix}

\end{document}